\documentclass[aps,twocolumn,showpacs,preprintnumbers,amsmath,amssymb,nofootinbib,superscriptaddress,showkeys]{revtex4}

\usepackage{hyperref}
\usepackage{float}
\usepackage{epsfig}
\usepackage{graphicx}
\usepackage{mathptmx}      % use Times fonts if available on your TeX system
\usepackage{latexsym}
\usepackage{longtable}
\usepackage{dcolumn}
\usepackage{natbib}
\usepackage{comment}
\newcommand{\EP}{\epsilon}

\begin{document}

\title{Low energy chiral two pion exchange potential with statistical
  uncertainties}

\author{R. Navarro P\'erez}\email{rnavarrop@ugr.es}
\affiliation{Departamento de F\'{\i}sica At\'omica, Molecular y
  Nuclear and Instituto Carlos I de F{\'\i}sica Te\'orica y
  Computacional \\ Universidad de Granada, E-18071 Granada, Spain.}
\affiliation{Department of Physics and Astronomy, Iowa State
  University, Ames, Iowa 50011, USA.}

\author{J.E. Amaro}\email{amaro@ugr.es} \affiliation{Departamento de
  F\'{\i}sica At\'omica, Molecular y Nuclear  and Instituto Carlos I
  de F{\'\i}sica Te\'orica y Computacional \\ Universidad de Granada,
  E-18071 Granada, Spain.}
  
\author{E. Ruiz
  Arriola}\email{earriola@ugr.es} \affiliation{Departamento de
  F\'{\i}sica At\'omica, Molecular y Nuclear  and Instituto Carlos I
  de F{\'\i}sica Te\'orica y Computacional \\ Universidad de Granada,
  E-18071 Granada, Spain.} 

\date{\today}

\begin{abstract} 
\rule{0ex}{3ex} We present a new phenomenological Nucleon-Nucleon
chiral potential fitted to 925 pp and 1743 np scattering data selected
from the Granada-2013 NN-database up to a laboratory energy of $125$
MeV with 20 short distance parameters and three chiral constants
$c_1$, $c_3$ and $c_4$ with $\chi^2/\nu = 1.02$.  Special attention is
given to testing the normality of the residuals which allows for a
sound propagation of statistical errors from the experimental data to
the potential parameters, phase-shifts, scattering amplitudes and
counter-terms. This fit allows for a new determination of the chiral
constants $c_1$, $c_3$ and $c_4$ compatible with previous
determinations from NN data. This new interactions is found to be
softer than other high quality potentials by undertaking a Weinberg
eigenvalue analysis. We further explore the interplay between the
error analysis and the assumed form of the short distance
interaction. The present work shows that it is possible to fit NN
scattering with a TPE chiral potential fulfilling all necessary
statistical requirements up to 125 MeV and shows unequivocal
non-vanishing D-wave short distance pieces.
\end{abstract}

\maketitle

%\section{Introduction}

\section{Introduction}

The NN interaction is beyond any doubt a key building block of nuclear
physics but, what are the decisive features which make the interaction
qualify for an {\it ab initio} description of binding in nuclei ?.
While there may not be a correct answer for this question, we will
provide what we think are important ingredients towards this goal. 

From a statistical point of view, a traditional figure of merit has
been the value of $\chi^ 2 /\nu $ after a least squares minimization
fit np and pp scattering data with $\nu=N-P$, the difference between
the number of fitted data and the number of fitting parameters. This
approach was initiated in 1957 at about pion production threshold
energies~\cite{PhysRev.105.302} (see
\cite{arndt1966chi,Machleidt:1992uz} for reviews) and extended up to
$3 {\rm GeV}$ more recently with a $\chi^2/\nu \sim
1.4$~\cite{Arndt:2007qn}.  After the benchmarking Partial Wave
Analysis (PWA) of the Nijmegen group 20 years
ago~\cite{Stoks:1993tb,Stoks:1994wp}, the path and key necessary
features were shown to provide a statistically satisfactory fit,
i.e. having an expected $\chi_{\rm min}^2 /\nu \sim 1 \pm
\sqrt{2/\nu}$ within $1\sigma$ confidence level up to about pion
production threshold: suitable data selection, incorporating charge
dependent One Pion Exchange (OPE) interactions as well as many EM
features and an adequate statistical interpretation of
results~\cite{Bergervoet:1988zz,Stoks:1992ja}. This
$\chi^2/\nu$-values have set the standards in high quality NN
studies~\cite{Stoks:1994wp,Wiringa:1994wb,Machleidt:2000ge,Gross:2008ps,Perez:2013mwa,
  Perez:2013jpa,Perez:2013oba,Perez:2014yla}.  The least squares
$\chi^2$-fit approach uses selected experimental data with
uncertainties which should be described in terms of a postulated
theory according to accepted statistical principles. In particular, if
the theory is flexible enough the difference between the actual
measured data and the proposed theory should be a statistical
fluctuation. The size of the fluctuation is controlled by the number
of available data as well the reported experimental uncertainties.
This is the essence of the normality test for residuals which
relevance we have recently stressed~\cite{Perez:2014yla,Perez:2014kpa}
(see Refs.~\cite{Bergervoet:1988zz,Stoks:1992ja} for related ideas).
The main advantage is that if this test is passed correctly we expect
the addition of new data in the future to sharpen the estimates of the
theoretical parameters.

The standard choice of pion-production threshold as upper limit at CM
momentum $p \sim \sqrt{M_N m_\pi} \sim 360 {\rm MeV}$ is essentially
based on the simplicity of treatment, as one may ignore the explicit
contribution of the inelastic $NN \to NN\pi$ channel, but it does not
tell anything on the shortest physical length scale operating in the
binding of finite nuclei. Fortunately, even for nuclear matter
characterized by the Fermi momentum $p_F \sim 250 {\rm MeV}$ the role
of these inelasticities is negligible since $p_F \lesssim \sqrt{M_N
  m_\pi}$, and thus one may reduce the upper fitting energy, the more
the lighter the nucleus. Afnan and Tang recognized this for the case
of $^3$He and $^4$He~\cite{Afnan:1968zj} where good binding energies
were achieved when S-waves are fitted up to $E_{\rm LAB} \lesssim 100
{\rm MeV}$. Using simple coarse grained interactions and mean field
wave functions we have verified this feature for nuclei as heavy as
$^{40}$Ca~\cite{NavarroPerez:2011fm}.

On the hadronic scale, finite nuclei are weakly bound objects of
neutrons and protons and thus their de Broglie wavelength is large
enough for them to behave effectively as elementary particles. On the
other hand, when nucleons are far appart, say $r \gtrsim 2 \, {\rm
  fm}$, they do not to overlap and their interaction resembles a van
der Waals type of exchange of pions between point like nucleons (see
e.g. \cite{RuizArriola:2009vp,Cordon:2011yd} for a quark cluster point
of view). In such a case the corresponding scattering partial wave
amplitudes containing $n-\pi$ exchanges are analytical in the complex
CM-momentum plane with branch cuts at $p = \pm i n m_\pi/2$. This
provides upper limits in the maximal energy on the number of exchanged
pions which should be taken into account to represent the scattering
amplitude with the correct analytical structure. At pion production
threshold this gives $n \sim 2 \sqrt{M_N/m_\pi} \sim 5 $ pion
exchanges, which seems almost impossible to implement. In practice,
the strength of the discontinuity of the scattering amplitude may be
small enough to relax this requirement.

From the point of view of Quantum Chromo-Dynamics (QCD) hadronic
interactions can be described with sub-nuclear degrees of freedom like
quarks and gluons and lattice calculations for NN potentials have been
pursued in terms of these fundamental degrees of
freedom~\cite{Aoki:2011ep, Aoki:2012tk}. On the other hand, the
spontaneous breaking of chiral symmetry allows to derive a NN
interaction with multiple pion exchange for the long range part in
terms of an effective low energy Lagrangian where pions enter via
derivative couplings of the field $\sim \partial \Phi/f_\pi \sim p
/f_\pi \Phi$ with $f_\pi = 92 {\rm MeV}$ the pion weak decay constant
~\cite{Weinberg:1990rz, Ordonez:1993tn,Kaiser:1997mw}. Actually, the
breakdown scale $\Lambda_\chi$ for a theory with just pions and
nucleons is expected to be at the branch cut $p= \pm i m_R/2$
corresponding to pionic resonance exchanges of mass $m_R$, hence
$\Lambda_\chi \sim |p|= m_R/2 $.  A large $N_c$ quark-hadron duality
argument gives $m_\rho \sim \sqrt{ 24 \pi/N_c}
f_\pi$~\cite{Masjuan:2012sk} and more complete treatments yield indeed
similar estimates for $m_R$ with
$R=\rho,A_1,\pi^*,\sigma$~\cite{Ledwig:2014cla}. Therefore the
strength of the discontinuity of the scattering amplitude due to
chiral $n \pi$ exchange is suppressed as $(m_\pi
n/(2\Lambda_\chi))^{2n}$.  Thus, from the point of view of the
operationally {\it needed} and the theoretically {\it available}
higher scale the consideration of chiral TPE seems like a perfect
match up to $p \lesssim 3m_\pi/2$ corresponding to the location of
$3\pi-$exchange cut and a LAB energy of $T_{\rm LAB} \sim 90 {\rm
  MeV}$.

This type of chiral effective interactions can be implemented as
standard quantum mechanical potentials expanded in powers of momentum
relative to $\Lambda_\chi$ and still require the use of
phenomenological counter-terms featuring the integrated out short
distance behavior. Thus, a comparison of chiral potentials to NN
scattering data is indispensable, even at very low
energies~\cite{Rentmeester:1999vw, Entem:2003ft, Epelbaum:2004fk,
  Machleidt:2011zz}. Since the mid-nineties several interactions, at
different orders, have been developed attempting to accurately
describe NN scattering processes with chiral components as their main
feature. In recent years a trend to compromise the description of
intermediate and high energy data in exchange for a more accurate
representation of low energy data, $E_{\rm LAB} \lesssim 125 {\rm
  MeV}$, has
emerged~\cite{Ekstrom:2013kea,Gezerlis:2013ipa,Ekstrom:2014dxa,
  Gezerlis:2014zia}. The non-trivial question is whether this
theoretical expectation is confirmed by the statistical analysis of
the currently available data below those energies.

Moreover, this reduction in the fitted energy range implies a
trade-off between improved theoretical reliability and a loss of many
scattering data in the analysis. This may also imply a loss of
precision and, as a consequence, a loss of predictive
power~\cite{Amaro:2013zka}. This paper studies this interplay between 
precision and predictive power by fitting a  chiral potential
to  low energy data, $E_{\rm LAB} \lesssim 125 {\rm
  MeV}$ and undertaking the statistical uncertainties.  

Using a Delta-Shell (DS) potential initially proposed by
Avil\'es~\cite{Aviles:1973ee} and rediscovered in
Ref.~\cite{Entem:2007jg} it was possible to coarse grain the NN
interaction by proving it at certain sensible
points~\cite{NavarroPerez:2011fm, NavarroPerez:2012qf}. To select a
self-consistent data base of over $6700$ scattering data up to
laboratory energy of $350$MeV we fitted a DS potential with a one pion
exchange (OPE) potential tail starting at $3.0$ fm and electromagnetic
EM effects~\cite{Perez:2013mwa, Perez:2013jpa}. Once the data base was
fixed we modified the DS potential including a chiral two pion
exchange $\chi$TPE tail starting at $1.8$fm and made a new
determination of the chiral constants $c_1$, $c_3$ and $c_4$ with
statistical uncertainties~\cite{Perez:2013oba}. Also a local and
smooth potential, that describes the same
database~\cite{Perez:2014yla}, has allowed to propagate statistical
uncertainties into few body calculations~\cite{Perez:2014laa}. The
basic requirement of normally distributed data for any least squares
fit is verified; it has been checked for all three phenomenological
potentials previously mentioned~\cite{Perez:2014yla}.

The idea of coarse graining in Nuclear Physics pioneered by
Avil\'es~\cite{Aviles:1973ee}, has a modern correspondence in the
$V_{\rm lowk}$ approach~\cite{Bogner:2003wn} or the Similarity
Renormalization Group (SRG)~\cite{Bogner:2006pc} (see also
\cite{Anderson:2008mu}) implemented over a decade ago for Nuclear
Structure calculations.  That approach allows to take advantage of the
universal character of the NN interaction from a Wilsonian viewpoint
while keeping the scattering amplitude unchanged and conveniently
softening the repulsive core. However, this requires a basic starting
bare interaction which, as alreay mentioned inherits both statistical
and systematic uncertainties from the scattering data. The propagation
of these uncertainties in the SRG scheme becomes computationally
costly, as any fluctuation of the interaction would require a new SRG
analysis.
 
One of the main advantages of using a coarse grained interaction,
whether it is a $V_{\rm lowk}$ potential, an SRG evolved interaction,
an oscillator basis representation or a delta-shell potential, is the
intrinsic softening of the short range part as compared to other
realistic interactions fitted directly to scattering NN data up to
pion production threshold. However, the softness of the interaction
can be quantitatively determined, and therefore compared, by finding
the largest Weinberg eigenvalue~\cite{Weinberg:1963zza}. In fact, this
type of analysis provides useful information on the convergence rate
(if any) of the Born series.

%\section{Low energy potential}

In this work we present a new DS-$\chi$TPE potential fitted to low
energy data up to $125$MeV LAB energy. This has practical advantages
as the core gets reduced improving on the suitability of mean field
schemes \cite{NavarroPerez:2011fm} because the effective interaction
evolves with this upper fitted energy~\cite{NavarroPerez:2013iwa}. We
have also previously reported on the consequences of reducing the
upper limit~\cite{Amaro:2013zka} and how the statistical uncertainties
of phase-shifts and shell model matrix elements increase to the point
of making OPE and $\chi$TPE indistinguishable. This would be a
situation where the only advantage of $\chi$TPE over OPE would be in
the reduction of the number of parameters, but not so much in a better
quality in the description of the data.

Finally, we hasten to emphasize that ours {\it is not} a conventional
$\chi$PT calculation; we use long range potentials above a certain
distance $r_c$ and coarse grain the short part of the interaction
below that distance with a sampling $\Delta r \sim \hbar /p_{\rm max}$
resolution~\cite{Perez:2013cza}, but take no position on how the short
distance piece should be organized within a perturbative setup. In
this regard, let us mention that while there is agreement on the long
distance features of multi-pionic exchange interactions based on
$\chi$PT, much has been said on the way the short distance pieces of
the interaction should be organized. The discussion on the specific
power counting to be applied within $\chi$PT has been around since the
very beginning and most discussions have been carried out on the basis
of theoretical consistency~\cite{Nogga:2005hy, PavonValderrama:2005wv,
  Birse:2005um, Epelbaum:2006pt, Valderrama:2009ei, Valderrama:2011mv}
(a comprehensive discussion was provided also in
\cite{Machleidt:2011zz} and references therein).  To date these
alternative schemes have not been seriously confronted to experimental
$np$ and $pp$ scattering data {\it directly} as we do here by using
the classical statistical $\chi^2$ least squares approach. Of course,
a proper discussion on power counting requires an {\it a priori}
assessment on the expected size of the neglected counterterms. We will
leave this discussion for future work where the Bayesian viewpoint
proposed in Ref.~\cite{Furnstahl:2014xsa} looks promising.

The paper is organized as follows. In Section \ref{sec:fit} we present
our potential and the necessary details for the fit, the normality
issues, error propagation and a study of the potential softness via
the Weinberg eigenvalues analysis. In Section \ref{sec:scatt} we
generate scattering properties including phase shifts, scattering
amplitudes and low energy threshold parameters. The low momentum
structure of the theory is presented and discussed in Section
\ref{sec:eff-int}. In Section \ref{sec:comparison} we analyze other
existing approaches in the literature and discuss in detail their
statistical features.  Finally, in Section \ref{sec:concl} we come to
our main conclusions.

\section{Coarse grained potential}
\label{sec:fit}

For a motivation on the use of a coarse grained potential in nuclear
physics we refer to Ref.~\cite{NavarroPerez:2011fm}. The anatomy of
the NN potential including multi-pion exchange and the expected number
of fitted parameters has been discussed in Ref.~\cite{Perez:2013cza}.

\subsection{Form of the potential}

The structure of the potential is the same as the DS-$\chi$TPE
potential of~\cite{Perez:2013oba} with a clear boundary $r_c=1.8$fm
between the short range phenomenological part with delta-shells and
the long range tail with one and two pion exchange plus
electromagnetic interactions.
\begin{equation}
 V(r) = V_{\rm DS}(r) + [V_{\rm OPE}(r) + V_{\rm TPE}(r) + V_{\rm EM}]
\theta(r-r_c).
\label{eq:potential}
\end{equation}
The long range potentials are the same as the ones used
in~\cite{Perez:2013oba}, although the chiral constants $c_1$, $c_3$ and
$c_4$ in $V_{\rm TPE}(r)$ are used as fitting parameters. The DS part is
given by
\begin{equation}
 V_{\rm DS} = \sum_{n=1}^{21} O_n \left[\sum_{i=1}^N V_{i,n} \Delta r
\delta(r-r_i) \right], \ \ r \leq r_c
\end{equation}
where $O_n$ is the set of operators in the extended AV18 basis detailed
in Appendix A of~\cite{Perez:2013jpa}, $V_i,n$ are strength
coefficients, $r_i$ are the concentration radii and $\Delta r = 0.6$fm
is the distance between them. As with previous works we decompose the
potential into partial waves by 
\begin{equation}
 V_{l,l'}^{J,S}(r) = \frac{1}{2\mu_{\alpha\beta}}\sum_{i=1}^N
 (\lambda_i)_{l,l'}^{S,J}\delta(r-r_i), \ \ r \leq r_c
\end{equation}
and use the 15 lowest angular momentum partial waves to parameterize the
full potential and calculate the more peripheral partial waves by
decomposing back from the operator basis to the partial wave basis.

\begin{figure*}
\centering
\includegraphics[width=\linewidth]{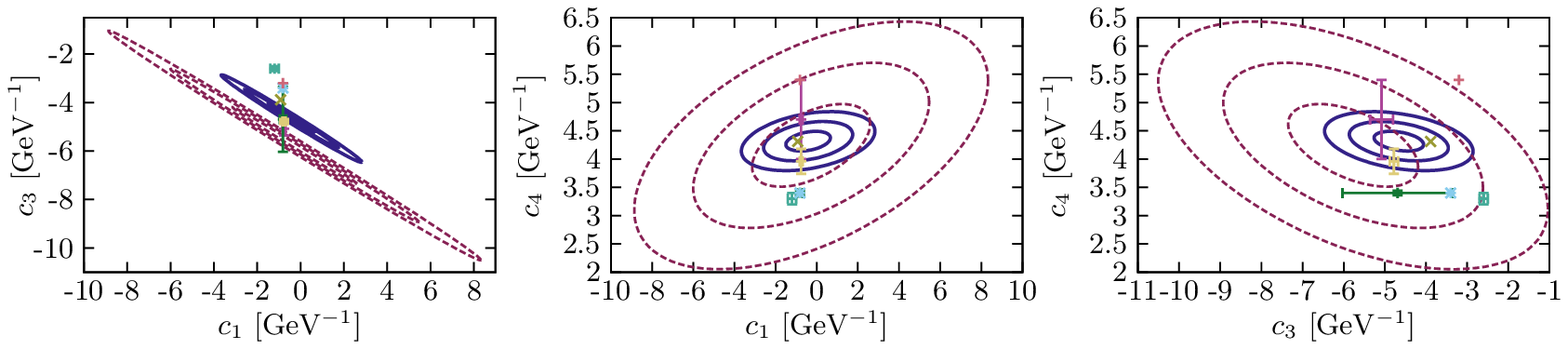}
\caption{(Color online) Correlation ellipses of the chiral constants
$c_1$, $c_3$ and $c_4$ determined from a fit to the self consistent
database of~\cite{Perez:2013jpa} with a DS-$\chi$TPE potential including
data with $T_{\rm LAB} \leq 350$~\cite{Perez:2013oba} (blue solid lines)
and $T_{\rm LAB} \leq 125$ (red dashed lines). The concentric ellipses
give, from the smallest to the largest one, the $68\%$, $95\%$ and
$99\%$ confidence regions respectively. The points and crosses
correspond to the determinations listed in Table VI
of~\cite{Perez:2013oba}} 
\label{fig:LECellipse}       % Give a unique label
\end{figure*}

The choice above is based on the high quality of our previous fit up
to $T_{\rm LAB} = 350 {\rm MeV}$ which fulfills all needed
statistical tests (see below). Our aim is that by reducing the energy
range to $T_{\rm LAB} = 125 {\rm MeV}$ correlations among the
parameters will appear implying a reduction of the number of
independent parameters. Although the error of the data below $T_{\rm
  LAB} = 125 {\rm MeV}$ is the same, their induced propagation is
amplified as a result of a fitting a smaller number of data. Even the
observables computed below $125 {\rm MeV}$ exhibit a larger error.

\subsection{Fitting short distance parameters}

Below $T_{\rm LAB} \leq 125$MeV the self-consistent data base obtained
in~\cite{Perez:2013jpa} contains $N_{pp}=925$ pp data and
$N_{np}=1743$ np data including normalizations. This upper limit on
the laboratory frame energy allows to reduce the number of parameters
from $30$ in~\cite{Perez:2013oba} to $20$, appart from the $3$ chiral
constants. Of course, with less data constraining the interaction the
statistical uncertainties in the potential parameters are larger. The
resulting delta-shell fitting parameters yield a total value of
$\chi^2/\nu=1.02$ and are shown in Table~\ref{tab:LEDeltaShellsTPE}.
Note that to $1\sigma$ confidence level, one expects $\chi^2/\nu = 1
\pm \sqrt{2/\nu}$, which in this particular case means $0.097 \le
\chi^2/\nu \le 1.03$.

\begin{table}
 \caption{\label{tab:LEDeltaShellsTPE} Delta-shell parameters fitted to
reproduce 2668 pp and np scattering data with $T_{\rm LAB} \leq 125$MeV.
Statistical error bars are propagated from experimental uncertainties.
The complete potential has a $\chi$TPE tail for $r > 1.8$fm and all
relevant electromagnetic interactions.}
 \begin{ruledtabular}
 \begin{tabular*}{\columnwidth}{@{\extracolsep{\fill}} l D{.}{.}{1.5} 
 D{.}{.}{2.5}  D{.}{.}{2.6}  }
 Wave  & \multicolumn{1}{c}{$\lambda_1$} 
       & \multicolumn{1}{c}{$\lambda_2$} 
       & \multicolumn{1}{c}{$\lambda_3$} \\
       & \multicolumn{1}{c}{($r_1=0.6$fm)} 
       & \multicolumn{1}{c}{($r_1=1.2$fm)} 
       & \multicolumn{1}{c}{($r_1=1.8$fm)} \\
 \hline \noalign{\smallskip}
 $^1S_0{\rm np}$&  0.88(79) & -0.75(19) & -0.053(46) \\ 
 $^1S_0{\rm pp}$&  1.9(2)   & -0.89(3)  & -0.028(9)  \\ 
 $^3P_0$        &     $-$   &  0.40(13) & -0.061(26) \\ 
 $^1P_1$        &     $-$   &  1.06(9)  &     $-$    \\ 
 $^3P_1$        &     $-$   &  1.5(1)   &  0.016(13) \\ 
 $^3S_1$        &  1.6(6)   &     $-$   &     $-$    \\ 
 $\varepsilon_1$&     $-$   & -2.78(8)  & -0.19(4)   \\ 
 $^3D_1$        &     $-$   &  3.0(5)   &     $-$    \\ 
 $^1D_2$        &     $-$   & -0.58(7)  &     $-$    \\ 
 $^3D_2$        &     $-$   &     $-$   & -0.28(1)   \\ 
 $^3P_2$        &     $-$   & -0.44(1)  &     $-$    \\
 $\varepsilon_2$&     $-$   &     $-$   &  0.097(11) \\ 
 $^3F_2$        &     $-$   &     $-$   &     $-$    \\
 $^1F_3$        &     $-$   &     $-$   &     $-$    \\ 
 $^3D_3$        &     $-$   &  1.1(1)   &     $-$    \\ 
 \end{tabular*}
 \end{ruledtabular}
\end{table}

This fit provides a new determination of the chiral constants
$c_1=-0.27 \pm 2.87$, $c_3 = -5.77 \pm 1.58$ and $c_4 = 4.24 \pm 0.73$
GeV$^{-1}$ which is mostly compatible to the one
from~\cite{Perez:2013oba}. In Figure~\ref{fig:LECellipse} we compare
the ellipses of the present fit with those of our previous fit to 350
MeV~\cite{Perez:2013oba}. Although each individual constant, with its
corresponding $1\sigma$ confidence interval, is statistically
compatible with the determination with data up to $T_{\rm LAB} = 350$
MeV, the correlation ellipses of both fits for the $c_1,c_3$
pair do not overlap.

\begin{figure}
\centering
\includegraphics[width=\linewidth]{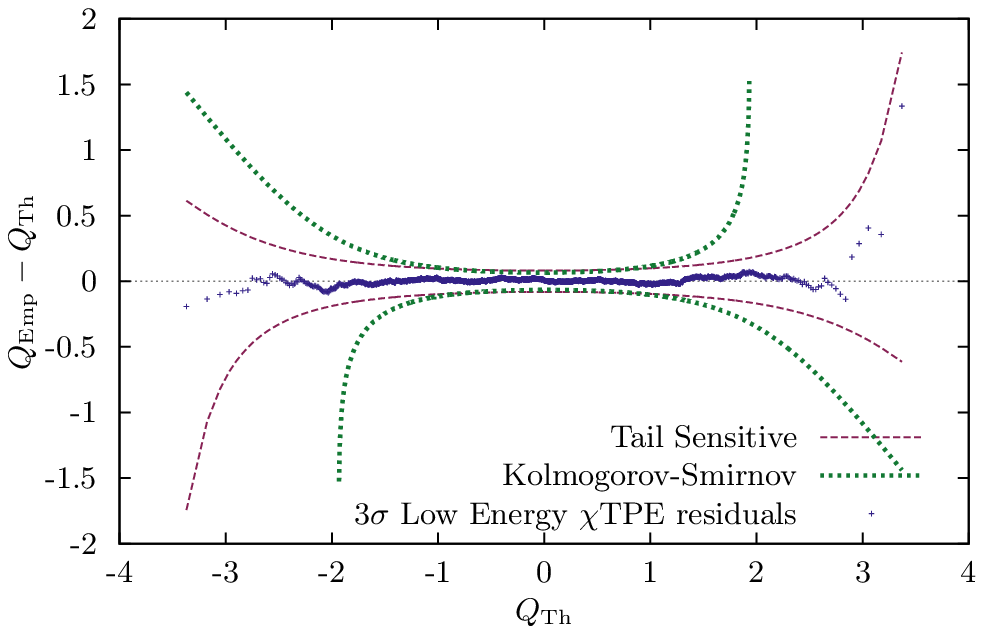}
\caption{(Color online) Rotated quantile-quantile plot for the
  residuals of the low energy DS-$\chi$TPE potential for $E_{\rm LAB}
  \le 125 {\rm MeV}$ (blue crosses). The Kolmogorov-Smirnov (dotted
  green line) and Tail sensitive (dashed red line) confidence bands
  with an $\alpha=0.05$ significance level are also included.}
\label{fig:qqplot}       % Give a unique label
\end{figure}  

\section{Scattering properties}
\label{sec:scatt}

\subsection{Normality test}

Once the potential parameters are fitted to the self-consistent data
base we check the normality of the residuals
\begin{equation}
 R_i = \frac{O_i^{\rm exp} - O_i^{\rm theo}}{\Delta O_i^{\rm exp}}.
\end{equation}
To this end we apply the recently introduced Tail-Sensitive (TS)
test~\cite{Aldor:2013} (see Refs.~\cite{Perez:2014yla,Perez:2014kpa}
for details and practical implementations). The TS test compares the
quantiles of an empirical distribution with the quantiles of a normal
distribution. The finite size of the sample gives a confidence
interval for each quantile which can be calculated analytically for a
previously determined significance level. If an empirical quantile
falls outside of its corresponding confidence interval the hypothesis
of normality is rejected. In figure~\ref{fig:qqplot} we show a rotated
quantile-quantile (QQ) plot comparing the theoretical quantiles of the
standard normal distribution $N(0,1)$ with the residuals quantiles;
the confidence intervals for the TS and the more familiar
Kolmogorov-Smirnov (KS) tests are also shown with a significance level
$\alpha=0.05$. As can be seen, the empirical residuals resulting for
the low energy fit always fall within the $95\%$ confidence bands of
both the TS and KS normality tests.

Aside from obtaining a graphical representation, testing for normality
is a straightforward procedure which simply requires to calculate a
quantity known as a test statistic $T$ and compare it with a
previously tabulated (or parameterized) critical value $T_c$ as a
function of the sample size $N$. Depending on the definition of $T$ on
each normality test, a larger (or conversely, smaller) $T$ indicates
larger deviations from the normal distribution and $T > T_c$ (or $T <
T_c$) gives significant evidence to reject the hypothesis of
normality; in the particular case of the TS test large deviations from
the normal distribution result in small values for $T$. For the TS
test a recipe for calculating $T$, a table of $T_c$ for $N \leq 50$
and a parameterization for $50 < N < 9000$ can be found
in~\cite{Perez:2014kpa}. The residuals of the low energy fit to the
self consistent database give $T=0.0068$ and the critical value for a
sample size $N=2668$ and a significance level $\alpha=0.05$ is $T_c =
0.0008$ and therefore there is not significant evidence to reject the
normality of the residuals.

\subsection{Error propagation}

Once the normality test is passed, we may proceed to propagate the
errors inherited by the theory through the $\chi^2$-fit. We do so
below for the phase shifts, the full scattering amplitude and the low
energy threshold parameters. Several schemes are
possible~\cite{Perez:2014jsa}: i) the standard covariance matrix of
building derivatives in quadrature with correlations, ii) the Monte
Carlo method based on a multivariate gaussian distribution based on
the $\chi^2$ function, and iii) the more elaborated bootstrap
method~\cite{Perez:2014jsa}. Results are fairly similar in all three
cases and we use for definiteness the method ii). It consists of
generating a sufficiently large sample drawn from a multivariate
normal probability distribution
\begin{equation}
\label{eq:multinormal}
 P(p_1,p_2,\ldots,p_P) = \frac{1}{\sqrt{(2 \pi)^P \det {\cal E}}}
 e^{-\frac{1}{2}({\bf p}- {\bf p}_0)^T {\cal E}^{-1} ({\bf p}- {\bf
 p}_0)},
\end{equation}
where ${\cal E}_{ij}= (\partial^2 \chi / \partial p_i \partial
p_j)^{-1}$ is the error matrix and ${\bf p}_0 $ are the fitting short
distance $\lambda$'s and chiral $c$'s minimizing the $\chi^2$. We
generate $M=1000$ samples ${\bf p}_\alpha \in P $ with $\alpha=1, \dots ,
M$, and compute $V_{NN}( {\bf \lambda_\alpha})$ from which the
corresponding observables are determined.

\begin{figure}
\centering
\includegraphics[width=\linewidth]{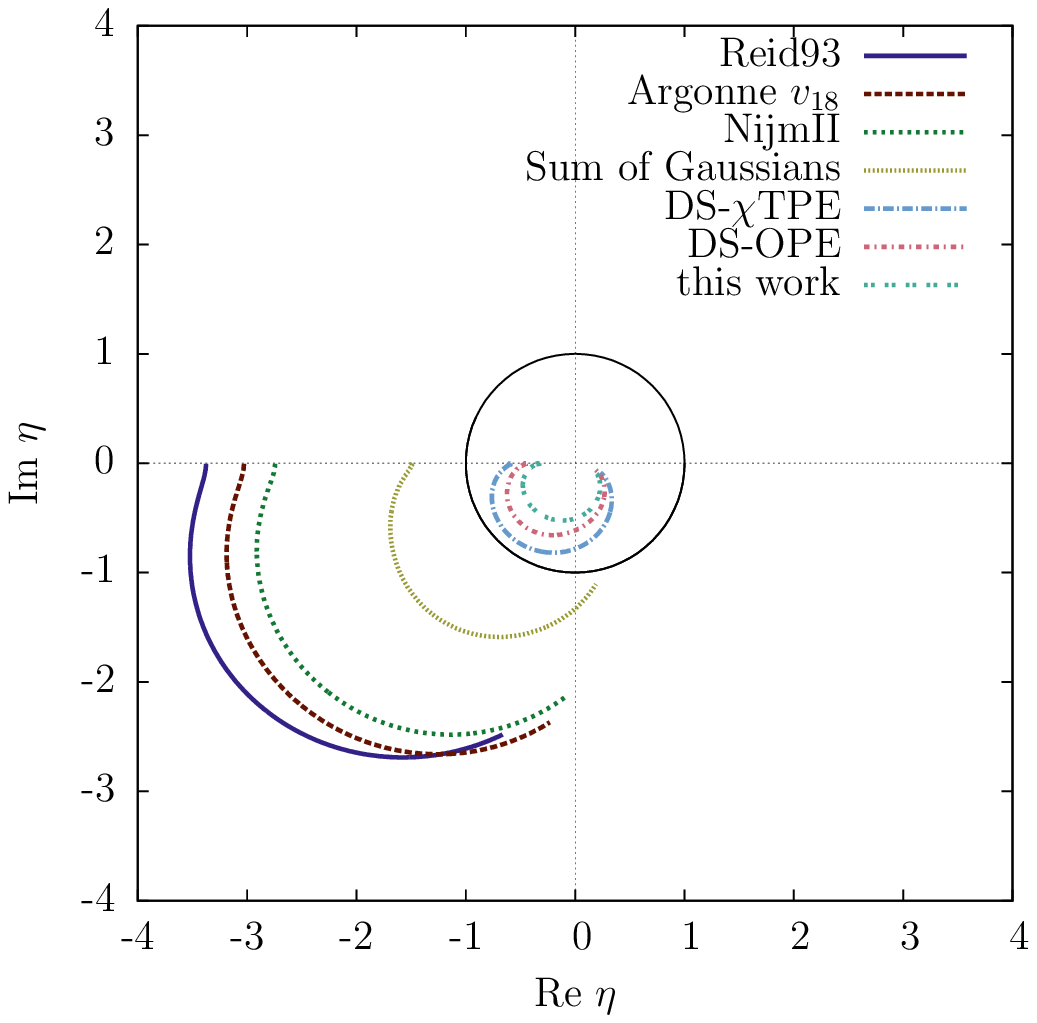}
\includegraphics[width=\linewidth]{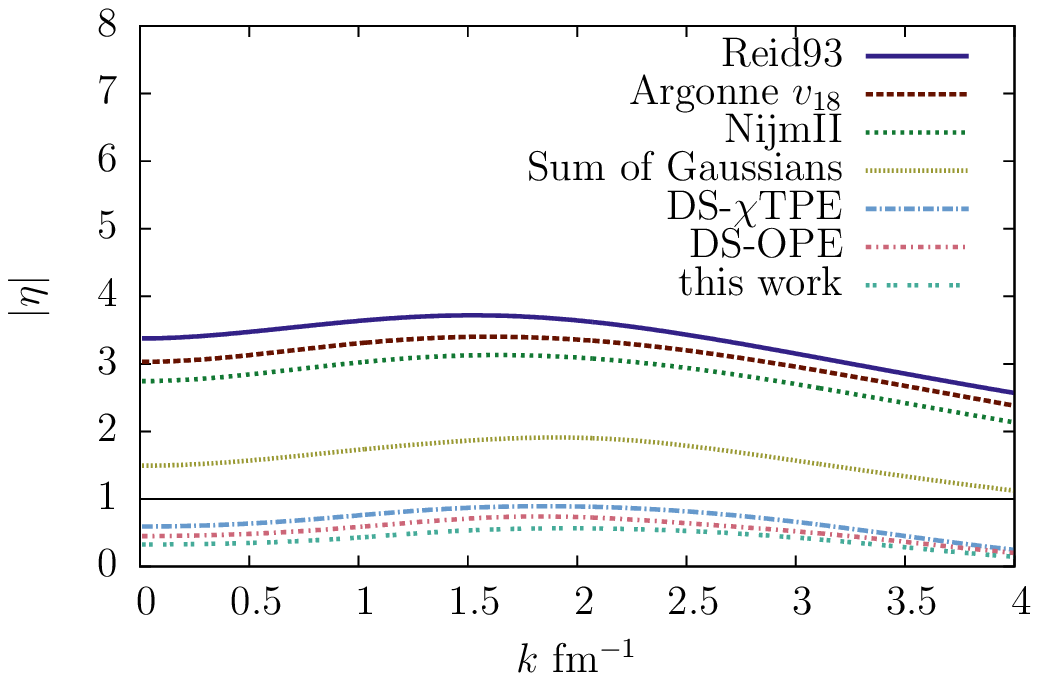}
\caption{(Color online) Trajectories (upper pannel) and modulus (lower
  pannel) of the largest repulsive Weinberg eigenvalue in the $^1S_0$
  channel as a funtion of CM momentum $k$ for high quality
  potentials. The potentials shown, ordered from largest to smalles
  modulus, are Reid93~\cite{Stoks:1994wp}, NijmII~\cite{Stoks:1994wp},
  AV18~\cite{Wiringa:1994wb}, Gauss-OPE~\cite{Perez:2014yla}
  DS-OPE~\cite{Perez:2013mwa,Perez:2013jpa},
  DS-$\chi$TPE~\cite{Perez:2013oba,Perez:2013cza}, and the present
  work. We show the $|\eta| = 1$ circle and line in a black solid line
  for reference }
\label{fig:weinberg-1S0}       % Give a unique label
\end{figure}

\subsection{Softness of the potential}

As we have mentioned in the introduction, restricting the fits to
lower maximal energies becomes less sensitive to short distance
details and hence one would expect satisfactory fits with softer cores
to be eligible. As can be seen from a comparison of
Table~\ref{tab:LEDeltaShellsTPE} with the corresponding one in the fit
to $350$MeV~\cite{Perez:2013oba}, the innermost strength coefficient
has larger uncertainties than the previous higher energy fit. Thus,
the softest edge of the confidence level could be used and still
describe the data. A more quantitative measure of the softness of the
potential is provided by the Weinberg eigenvalue
analysis~\cite{Weinberg:1963zza}. Within the $V_{\rm lowk}$
approach~\cite{Bogner:2003wn,Anderson:2008mu} this type of study
has shown that there is an effective softening of the strong
interaction by integrating out high CM momenta above $\Lambda \sim 2
{\rm fm}^{-1}$~\cite{Bogner:2006tw}. The analysis there was mainly
conducted in momentum space. So, we recapitulate the most relevant
aspects in configuration space for completeness. The main idea is to
solve the Schr\"odinger equation with a rescaled potential, $V \to
V/\eta$ by a complex parameter $\eta$. In the uncoupled case the
reduced wave equation reads
\begin{eqnarray}
-w_{l,\nu}'' (r) + \left[ \frac{l(l+1)}{r^2} w_{l,\nu} + \frac{U_l (r)}{\eta_{l,\nu} (k)}\right] w_{l,\nu}(r) = k^2 w_{l,\nu} (r)
\end{eqnarray}
with $U_l(r)= M V_l(r)$ the reduced potential. The solution is  
subject to the boundary conditions 
\begin{eqnarray}
w_{l,\nu} (r) &\sim& e^{i k r} \, , \qquad r \to \infty \nonumber  \\
w_{l,\nu} (r) &\sim& r^{l+1} \, , \qquad r \to 0 
\end{eqnarray}
corresponding to a regular solution at the origin and outgoing
spherical wave at infinity. On the upper lip of the positive energy
scattering cut, $ k \equiv \sqrt{k^2 + i 0^+}$ and the outgoing wave
becomes the normalizable Weinberg eigenfunction $w_{l,\nu}(r)$
corresponding to the Weinberg eigenvalue $\eta_{l,\nu}(k)$. They are
non-degenerate and are usually ordered according to the sequence
$|\eta_{l,1}(k)| > |\eta_{l,2}(k)| > \dots $. In general
$\eta_{l,\nu}(k)$ is complex and every single eigenvalue describes a
trajectory in the complex plane as a function of the real momentum
$k$.  Clearly, for purely imaginary momentum $k= i \gamma $ with
$\gamma> 0$ one has an exponential fall-off $\sim e^{-\gamma r}$
typical of a bound state, such as, e.g., the deuteron whose energy is
given by $E_d = - \gamma_d^2/M $. Thus, for the deuteron we have $\eta
(i \gamma_d)=1$. The important result is that the Born series for a
given angular momentum $l$ and CM momentum $k$ converges if and only
all eigenvalues are inside the unit circle in the complex plane, i.e.,
in particular $ |\eta_{l,1} (k)| \equiv \max_\nu |\eta_{l,\nu}(k)| <
1$.  Finally, the size of $ |\eta_{l,1}(k)|$ provides quantitative
information on the convergence rate in perturbation theory of the
scattering amplitude and hence of the Born
series~\cite{Weinberg:1963zza}.

In Fig.~\ref{fig:weinberg-1S0} we show the trajectories and modulus as
a function of $k$ for the largest Weinberg eigenvalue in the $^1S_0$
channel and for the high quality potentials
NijmII~\cite{Stoks:1994wp}, Reid93~\cite{Stoks:1994wp},
AV18~\cite{Wiringa:1994wb} (shown for illustration) as well as our own
high quality analyses using DS-OPE~\cite{Perez:2013mwa,Perez:2013jpa},
DS-$\chi$TPE~\cite{Perez:2013oba,Perez:2013cza}, the
Gauss-OPE~\cite{Perez:2014yla} and the present work's potentials. As
we see, all the delta-shell potentials including OPE and TPE and the
one of the present work generate Weinberg eigenvalues which are {\it
  always} inside the unit circle. The figure also convincingly shows
that the coarse graining of the NN interaction via delta-shells
actually yields a more perturbative potential when the maximum fitting
energy range is lowered.

\subsection{Phase-shifts}

\begin{figure*}
\centering
\includegraphics[width=0.7\linewidth]{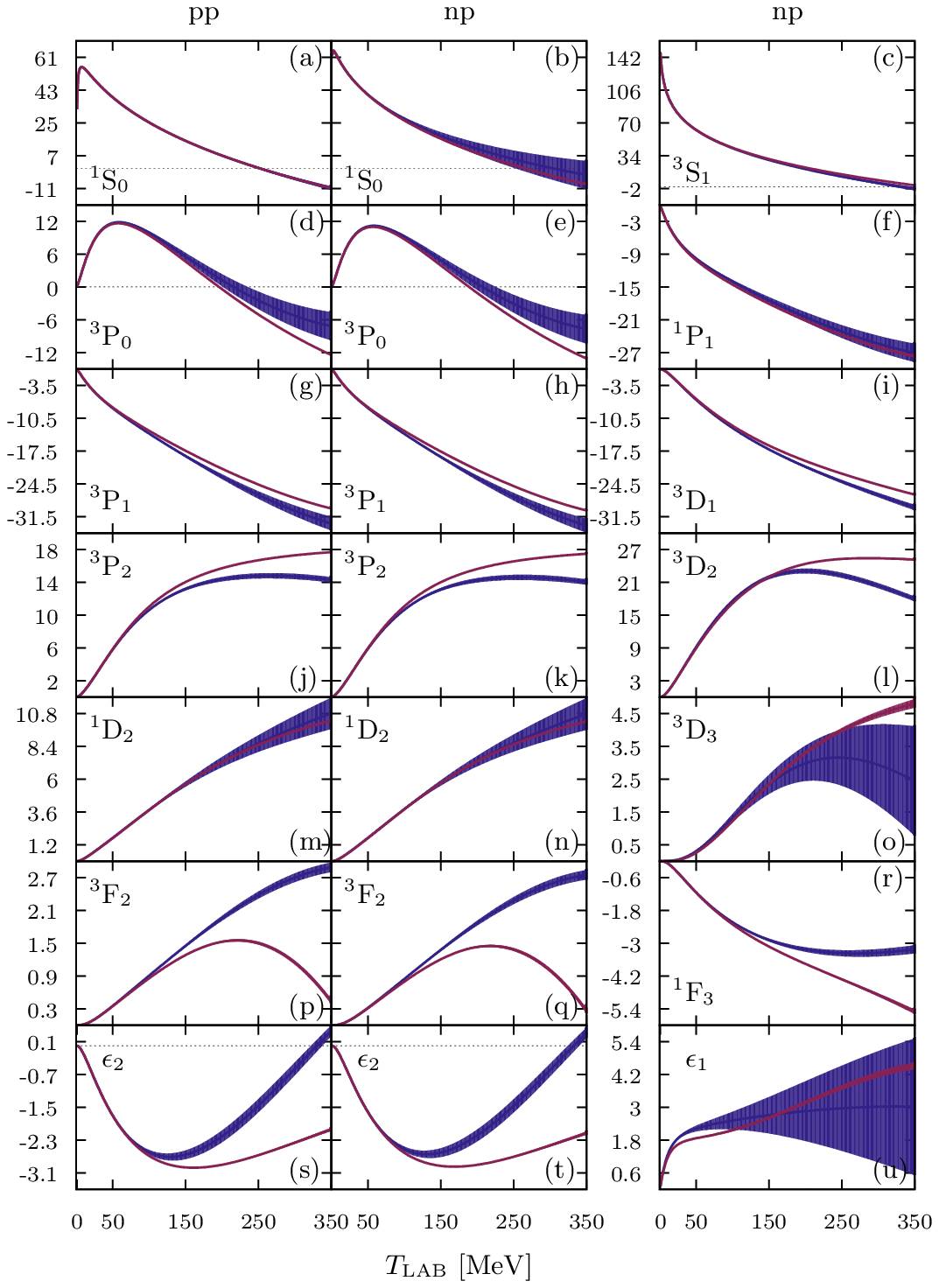}
\caption{(Color online) Phase-shifts for the low energy DS-$\chi$TPE
potential fitted to experimental data with $T_{\rm LAB} \leq 125$MeV
with statistical error bans (blue bands). The corresponding phase-shifts
and error bands for the full DS-$\chi$TPE reproducing data with $T_{\rm
LAB} \leq 350$ MeV~\cite{Perez:2013oba} are also included for comparison
(red bands)} 
\label{fig:Phaseshifts}       % Give a unique label
\end{figure*}

In Figure~\ref{fig:Phaseshifts} we show the low angular momentum
partial wave phase-shifts up to $T_{\rm LAB} = 350$MeV for the low
energy DS-$\chi$TPE and the DS-$\chi$TPE potential
of~\cite{Perez:2013oba}. The low energy version of the potential shows
larger statistical uncertainties at higher energies since there are no
data constraining the interaction above $125$ MeV. However, bellow
this energy value statistical error bands are also larger for the low
energy version of the potential. This indicates that the high energy
data also play a significant role in determining the uncertainties at
lower energies.

%\begin{comment}

\subsection{Wolfenstein parameters}

The full NN scattering amplitude reads
\begin{eqnarray}
M 
&=& a
+ m (\mathbf{\sigma}_1\cdot\mathbf{n})(\mathbf{\sigma}_2\cdot\mathbf{n}) 
+ (g-h)(\mathbf{\sigma}_1\cdot\mathbf{m})(\mathbf{\sigma}_2\cdot\mathbf{m}) 
\nonumber \\ 
&+& 
(g+h)(\mathbf{\sigma}_1\cdot\mathbf{l})(\mathbf{\sigma}_2\cdot\mathbf{l})  
+ c (\mathbf{\sigma}_1+\mathbf{\sigma}_2)\cdot\mathbf{n}
\label{eq:wolfenstein} 
\end{eqnarray}
where the Wolfenstein parameters $a,m,g,h,c$ depend on energy and
scattering angle, $\mathbf{\sigma}_1$ and $\mathbf{\sigma}_2$ are the
single-nucleon Pauli matrices, $\mathbf{l}$, $\mathbf{m}$,
$\mathbf{n}$ are three unitary orthogonal vectors along the directions
of $\mathbf{k}_f+\mathbf{k}_i$, $\mathbf{k}_f-\mathbf{k}_i$ and
$\mathbf{k}_i \wedge \mathbf{k}_f$, respectively, and ($\mathbf{k}_f$,
$\mathbf{k}_i$) are the final and initial relative nucleon
momenta. The relation with the phase shifts can be looked up in
Refs.~\cite{glockle1983quantum,Perez:2013jpa}.

\begin{figure*}
\centering
\includegraphics[width=0.8\linewidth]{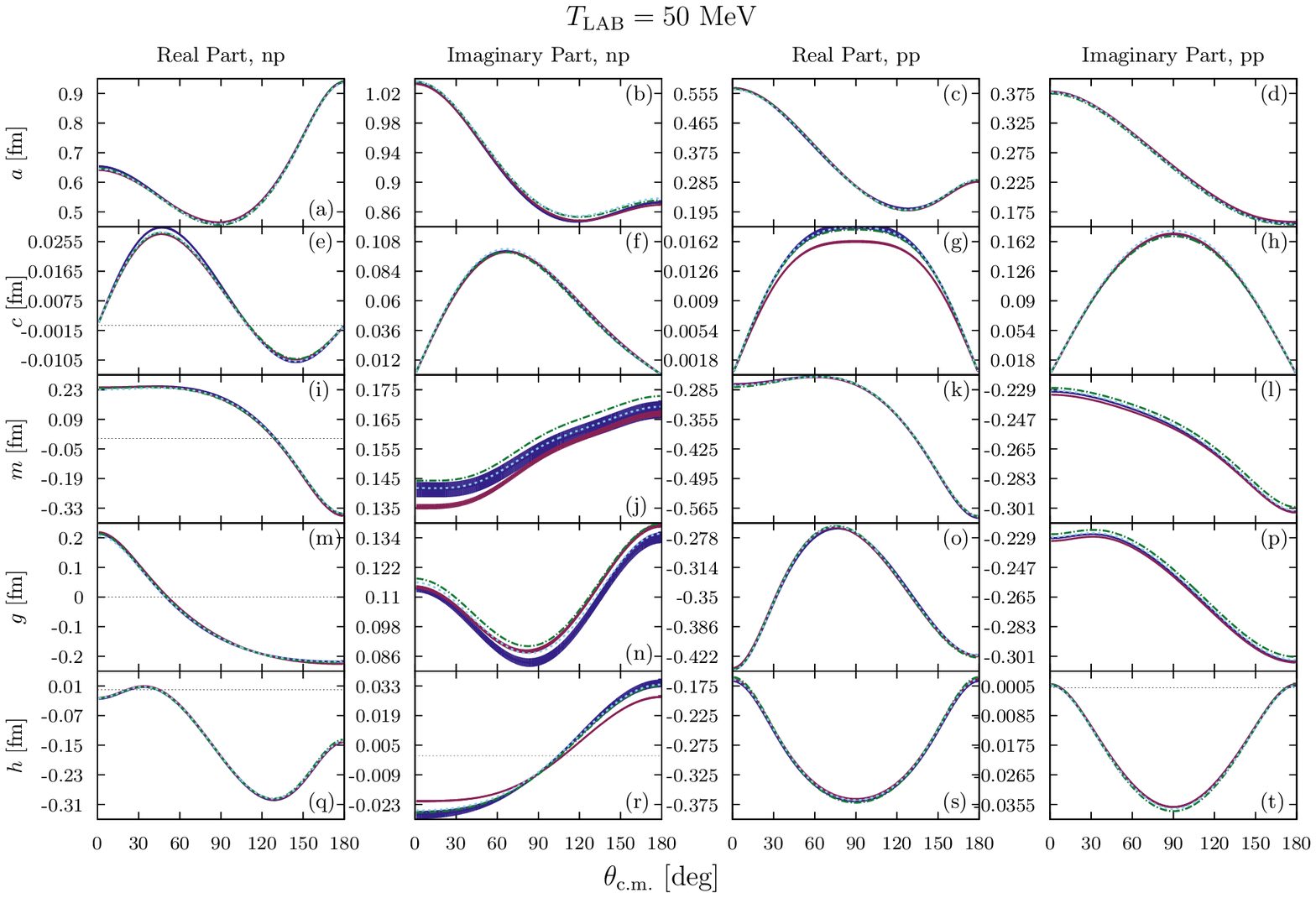}
\caption{(Color online) $np$ (left) and $pp$ (right) Wolfenstein parameters (in fm) as a function of the c.m. angle (in degrees) and for
$T_{\rm LAB} = 50$MeV. We compare the low energy DS-$\chi$TPE potential (blue band) with full DS-$\chi$TPE potential~\cite{Perez:2013oba} (red band) the Nijmegen PWA~\cite{Stoks:1993tb} (dotted, light blue line) and the AV18 potential~\cite{Wiringa:1994wb} (dashed-dotted, green line).} 
\label{fig:Wolfen050}       % Give a unique label
\end{figure*}

\begin{figure*}
\centering
\includegraphics[width=0.8\linewidth]{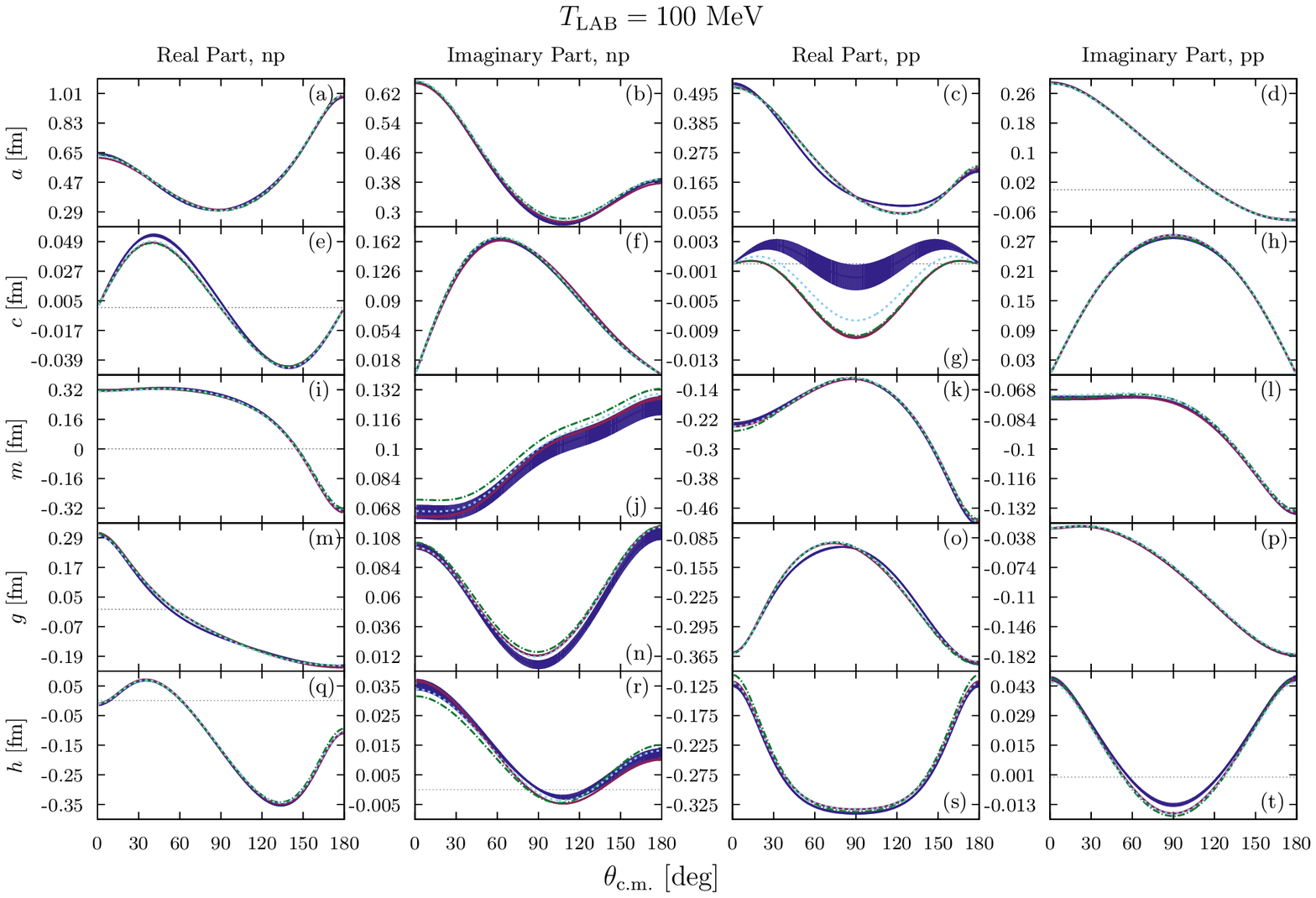}
\caption{(Color online) Same as Fig.~\ref{fig:Wolfen050} but for $T_{\rm LAB} = 100$MeV} 
\label{fig:Wolfen100}       % Give a unique label
\end{figure*}

\begin{figure*}
\centering
\includegraphics[width=0.8\linewidth]{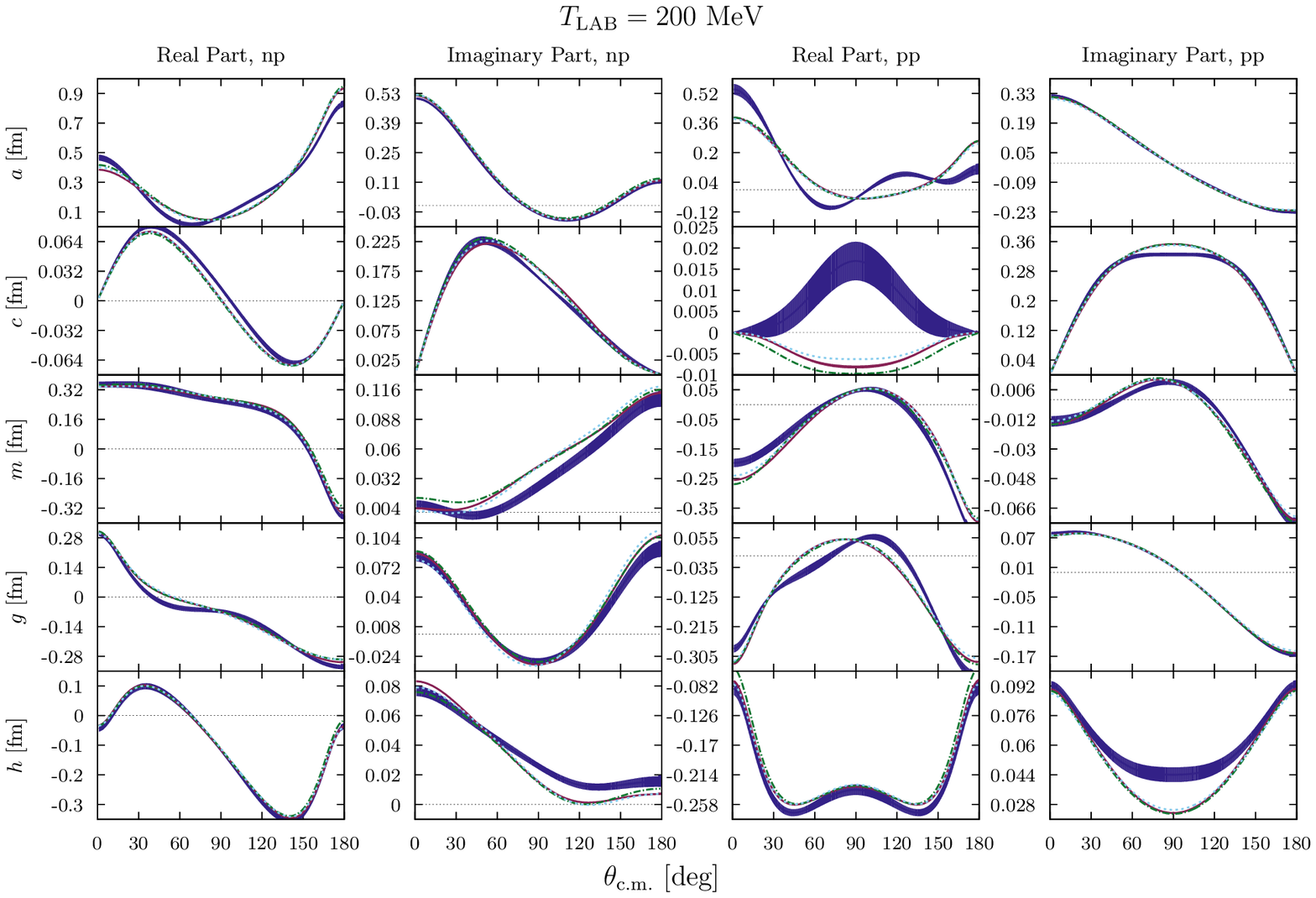}
\caption{(Color online) Same as Fig.~\ref{fig:Wolfen050} but for $T_{\rm LAB} = 200$MeV}  
\label{fig:Wolfen200}       % Give a unique label
\end{figure*}

Figures~\ref{fig:Wolfen050}, \ref{fig:Wolfen100}
and~\ref{fig:Wolfen200} compare the Wolfenstein parameters of the low
energy DS-$\chi$TPE potential with the full DS-$\chi$TPE ones at $50$,
$100$ and $200$ MeV respectively. At lower energies both interactions
show a fair level of agreement and again the low energy version of the
potential shows larger uncertainties. The discrepancies between both
potentials, indicated by non-overlapping error bands, can be
understood as the systematic uncertainties of both interactions. While
at $200$MeV the discrepancies are much larger and hence beyond the
range of validity of this new interaction, the dominance of systematic
over statistical errors has been a recurrent feature. Actually, this
worrysome pattern was pointed out in our early
studies~\cite{NavarroPerez:2012vr,Perez:2012kt} based on the
Nijmegen-1993 analysis and it reappears almost ubiquitously in any of
our own upgraded fits.

\subsection{Low energy threshold parameters}

In the absence of tensor force the phase-shifts with angular momentum
$l$ behave for low CM momentum, $k \to 0$, according to the effective
range expansion
\begin{eqnarray}
k^{2 l+ 1} \cot \delta_l(k)= - \frac1{\alpha_l}+ \frac12 r_l k^2 +
v_{2,l} k^4 + \dots \, ,
\end{eqnarray}
When the tensor force is considered we can apply a coupled channel
generalization of the effective range
expansion~\cite{PavonValderrama:2005ku}. Introducing the
$\mathbf{\hat{M}}$ matrix defined as
\begin{equation}
 \mathbf{DSD}^{-1} = \left(\mathbf{\hat{M}}+ik\mathbf{D}^2\right)\left(\mathbf{\hat{M}}-ik\mathbf{D}^2\right)^{-1},
\end{equation}
where $\mathbf{S}$ is the usual unitary S-matrix and $\mathbf{D}={\rm
  diag}(k^{l_1},\ldots,k^{l_N})$. In the limit $k \to 0$, the
$\mathbf{\hat{M}}$-matrix becomes
\begin{equation}
\label{MLowE}
 \mathbf{\hat{M}} = -\mathbf{a}^{-1} + \frac{1}{2} \mathbf{r} k^2 +
\mathbf{v_2} k^4 + \mathbf{v_3} k^6 + \mathbf{v_4} k^8 + \dots,
\end{equation}
where $\mathbf{a}$, $\mathbf{r}$ and $\mathbf{v_i}$ are the coupled
channel generalizations of $\alpha_0$, $r_0$ and $v_i$
respectively. We have recently evaluated them and confronted
statistical and systematic errors based on this
expansion~\cite{Perez:2014waa}.

Table~\ref{tab:LEPS-statistic} shows the low energy threshold
parameters of all partial waves with $j \leq 5$ for the low energy
DS-$\chi$TPE potential. The statistical uncertainties are propagated
by drawing $1000$ random sets of potential parameters following a
multivariate normal distribution according to the covariance matrix,
calculating the low energy threshold parameters for each set of
parameters and taking the mean and standard deviation.

\begin{table*}
 \caption{\label{tab:LEPS-statistic} Low energy threshold parameters for
all partial waves with $j \leq 5$ for the DS-$\chi$TPE potential. The
central value and statistical error bars correspond to the mean and
standard deviation of a population of $1000$ parameters calculated with
a Monte Carlo family of potential parameters drawn according the the
covariance matrix of the potential parameters. For each partial wave we
show the scattering length $\alpha$, the effective range $r_0$ and the
curvature parameters $v_2$, $v_3$ and $v_4$. For the coupled channels we
use the nuclear bar parameterization of the $S$ matrix. The units are in
powers of femtometers determined by the orbital angular momentum quantum
numbers $l$ and $l'$ of each partial wave}
 \begin{ruledtabular}
 \begin{tabular*}{\columnwidth}{@{\extracolsep{\fill}} c D{.}{.}{3.6}
D{.}{.}{3.6} D{.}{.}{3.6} D{.}{.}{3.6} D{.}{.}{3.6} }
 Wave   & \multicolumn{1}{c}{$\alpha({\rm fm}^{l+l'+1})$} 
        & \multicolumn{1}{c}{$ r_0({\rm fm}^{l+l'+1})$} 
        & \multicolumn{1}{c}{$ v_2({\rm fm}^{l+l'+3})$} 
        & \multicolumn{1}{c}{$ v_3({\rm fm}^{l+l'+5})$} 
        & \multicolumn{1}{c}{$ v_4({\rm fm}^{l+l'+7})$} \\
 \hline \noalign{\smallskip}
$^1S_0$ &   -23.739975 &     2.683075 &    -0.482309 &     3.876303 &   -19.536951 \\
        & \pm 0.019856 & \pm 0.010688 & \pm 0.010963 & \pm 0.043940 & \pm 0.122646 \\
$^3P_0$ &    -2.497645 &     3.809329 &     1.006610 &     3.861832 &    -7.889673 \\
        & \pm 0.007007 & \pm 0.017893 & \pm 0.013141 & \pm 0.044996 & \pm 0.083544 \\
$^1P_1$ &     2.780145 &    -6.461193 &    -1.707267 &     0.293702 &     7.970193 \\
        & \pm 0.008084 & \pm 0.029229 & \pm 0.050329 & \pm 0.094128 & \pm 0.083765 \\
$^3P_1$ &     1.514034 &    -8.667588 &     0.014268 &    -0.504544 &    -0.583288 \\
        & \pm 0.002833 & \pm 0.022467 & \pm 0.026836 & \pm 0.070704 & \pm 0.204727 \\
$^3S_1$ &     5.424721 &     1.833010 &    -0.120574 &     1.433963 &    -7.563664 \\
        & \pm 0.001887 & \pm 0.003057 & \pm 0.002884 & \pm 0.008492 & \pm 0.036475 \\
$\EP_1$ &     1.686135 &     0.426075 &    -0.243444 &     1.436825 &    -7.260537 \\
        & \pm 0.013563 & \pm 0.006913 & \pm 0.010099 & \pm 0.016457 & \pm 0.019671 \\
$^3D_1$ &     6.563492 &    -3.493365 &    -3.645026 &     1.135239 &    -2.638623 \\
        & \pm 0.026034 & \pm 0.013119 & \pm 0.021307 & \pm 0.017822 & \pm 0.020137 \\
$^1D_2$ &    -1.385117 &    14.895810 &    16.422846 &   -12.890261 &    37.903278 \\
        & \pm 0.002580 & \pm 0.045677 & \pm 0.118270 & \pm 0.193378 & \pm 0.601827 \\
$^3D_2$ &    -7.408744 &     2.851102 &     2.360846 &    -1.083437 &     1.753274 \\
        & \pm 0.005410 & \pm 0.003869 & \pm 0.012195 & \pm 0.026133 & \pm 0.026984 \\
$^3P_2$ &    -0.297390 &    -8.478863 &    -7.433750 &    -7.477923 &   -14.511322 \\
        & \pm 0.003170 & \pm 0.046753 & \pm 0.145525 & \pm 0.416712 & \pm 0.788810 \\
$\EP_2$ &     1.600387 &   -15.966142 &   -25.930128 &   -25.222196 &   -70.658514 \\
        & \pm 0.001190 & \pm 0.043977 & \pm 0.168760 & \pm 0.682425 & \pm 1.762290 \\
$^3F_2$ &    -0.974812 &    -5.957998 &   -24.133336 &   -83.161809 &  -124.079013 \\
        & \pm 0.001774 & \pm 0.058571 & \pm 0.209295 & \pm 0.899196 & \pm 3.081428 \\
$^1F_3$ &     8.377225 &    -3.925084 &    -9.873666 &   -15.298794 &    -2.050820 \\
        & \pm 0.002234 & \pm 0.001881 & \pm 0.010626 & \pm 0.056008 & \pm 0.129186 \\
$^3F_3$ &     2.680008 &   -10.037798 &   -20.952701 &   -20.389419 &   -30.275280 \\
        & \pm 0.001765 & \pm 0.011854 & \pm 0.056074 & \pm 0.257995 & \pm 0.643766 \\
$^3D_3$ &    -0.140354 &     1.371004 &     2.071825 &     1.913193 &    -0.549459 \\
        & \pm 0.002730 & \pm 0.000773 & \pm 0.003553 & \pm 0.013570 & \pm 0.019826 \\
$\EP_3$ &    -9.682379 &     3.260401 &     7.672521 &     9.579540 &    -1.135140 \\
        & \pm 0.000465 & \pm 0.000832 & \pm 0.004167 & \pm 0.019404 & \pm 0.047122 \\
$^3G_3$ &     4.874337 &    -0.030192 &     0.001640 &    -0.003425 &    -2.718540 \\
        & \pm 0.000639 & \pm 0.001672 & \pm 0.009409 & \pm 0.051735 & \pm 0.151814 \\
$^1G_4$ &    -3.212045 &    10.809086 &    34.473444 &    81.971319 &   104.049040 \\
        & \pm 0.000661 & \pm 0.003870 & \pm 0.024269 & \pm 0.155640 & \pm 0.521681 \\
$^3G_4$ &   -19.145092 &     2.058351 &     6.814736 &    16.772767 &    10.019825 \\
        & \pm 0.000743 & \pm 0.000143 & \pm 0.001002 & \pm 0.007057 & \pm 0.025122 \\
$^3F_4$ &    -0.016002 &    -3.053099 &    -4.815627 &    73.726022 &   664.426931 \\
        & \pm 0.001726 & \pm 0.002105 & \pm 0.012043 & \pm 0.055466 & \pm 0.126890 \\
$\EP_4$ &     3.585807 &    -9.548329 &   -37.136343 &  -185.113250 &  -587.360666 \\
        & \pm 0.000044 & \pm 0.002591 & \pm 0.014979 & \pm 0.061061 & \pm 0.413503 \\
$^3H_4$ &    -1.240294 &    -0.204717 &    -1.772049 &   -17.439098 &  -123.030299 \\
        & \pm 0.000290 & \pm 0.008206 & \pm 0.059293 & \pm 0.551235 & \pm 3.669909 \\
$^1H_5$ &    28.573515 &    -1.726914 &    -7.906396 &   -32.787619 &   -59.367511 \\
        & \pm 0.000317 & \pm 0.000034 & \pm 0.000320 & \pm 0.003254 & \pm 0.019300 \\
$^3H_5$ &     6.079919 &    -6.440909 &   -25.238708 &   -82.597219 &  -168.850963 \\
        & \pm 0.000281 & \pm 0.000514 & \pm 0.003801 & \pm 0.030382 & \pm 0.140866 \\
$^3G_5$ &    -0.009639 &     0.480549 &     1.878389 &     6.098743 &     6.785788 \\
        & \pm 0.000646 & \pm 0.000017 & \pm 0.000142 & \pm 0.001302 & \pm 0.006672 \\
$\EP_5$ &   -31.301936 &     1.556146 &     6.994315 &    28.175241 &    48.356412 \\
        & \pm 0.000033 & \pm 0.000020 & \pm 0.000180 & \pm 0.001704 & \pm 0.009264 \\
$^3I_5$ &    10.677985 &     0.010777 &     0.144456 &     1.427543 &     6.457572 \\
        & \pm 0.000110 & \pm 0.000058 & \pm 0.000542 & \pm 0.005504 & \pm 0.032654 \\
 \end{tabular*}
 \end{ruledtabular}
\end{table*}

\section{Effective interactions at low momentum}
\label{sec:eff-int}

We now turn to analyze the corresponding potential in momentum space,
particularly within a low momentum expansion. As we have shown
elsewhere~\cite{NavarroPerez:2013iwa}, the coefficients of the
expansion can be mapped into radial moments of volume integrals of the
potential, which exhibit some degree of universality. We will separate
the contributions stemming from the inner region $r < r_c$ containing
just delta-shell interactions and the outer region $r>r_c$ containing
the pion exchange potential tail. For ease of comparison, we will
consider the results in a Cartesian as well as in the spherical basis.

\subsection{Moshinsky-Skyrme parameters}

At the two body level the effective interaction of
Moshinsky~\cite{Moshinsky195819} and Skyrme~\cite{Skyrme:1959zz} can
be written as a pseudo-potential in the form
\begin{eqnarray} 
&& V_\Lambda ({\bf    p}',{\bf p}) 
=
 \int d^3 x e^{-i {\bf x}\cdot ({\bf p'}-{\bf p})}  \hat V({\bf x} ) 
 \nonumber \\ &&=  t_0 (1 + x_0 P_\sigma ) + \frac{t_1}2(1 + x_1
  P_\sigma ) ({\bf p}'^2 + {\bf p}^2) \nonumber \\ 
&&+  
 t_2 (1 + x_2
  P_\sigma ) {\bf p}' \cdot {\bf p} + 2 i W_0 {\bf S} \cdot({\bf p}'
  \wedge {\bf p}) \\ &&+ 
\frac{t_T}2 \left[ \sigma_1 \cdot {\bf p}
  \, \sigma_2 \cdot {\bf p}+ \sigma_1 \cdot {\bf p'} \, \sigma_2
  \cdot {\bf p'} - \frac13 \sigma_1 \, \cdot 
\sigma_2 ({\bf p'}^2+  {\bf p}^2)
\right] \nonumber \\  &&+
\frac{t_U}2 \left[ \sigma_1 \cdot {\bf p}
  \, \sigma_2 \cdot {\bf p}'+ \sigma_1 \cdot {\bf p'} \, \sigma_2
  \cdot {\bf p} - \frac23 \sigma_1 \, \cdot 
\sigma_2 {\bf p'}\cdot  {\bf p}
\right]  
+ {\cal O} (p^4) \nonumber 
\label{eq:skyrme2}
\end{eqnarray} 
where $P_\sigma = (1+ \sigma_1 \cdot \sigma_2)/2$ is the spin exchange
operator with $P_\sigma=-1$ for spin singlet ($S=0$), and $P_\sigma=1$
for spin triplet ($S=1$) states.

The effective interaction representation in terms of Moshinsky-Skyrme
parameters are presented in tables~\ref{tab:Skyrme}.  Since both
parameterizations consist of potential integrals, see
Ref.~\cite{NavarroPerez:2013iwa}, we show the contribution to the full
parameter by the phenomenological short range part $r \leq r_c$ and
the pion exchange tail $r> r_c$ with the corresponding
uncertainties. Since the potential is determined by low energy data
only, one would expect the short range contribution to counter-terms of
the most peripheral partial waves to be compatible with zero. However,
we see that although the errors are larger than those quoted
in~\cite{Perez:2014kpa} for a DS-$\chi$TPE potential fitted to data up
to $350$MeV, the short range counter-terms are never compatible with
zero. It is also worth noting that the full integrals both for
Moshinsky-Skyrme parameters and counter-terms show a large degree of
universality when compared to the same parameters for the DS-OPE and
DS-$\chi$TPE potentials shown in~\cite{Perez:2014kpa}.

\subsection{Counter-Terms}

The potential in momentum space can be written 
in the partial wave basis as 
\begin{eqnarray}
v^{JS}_{l',l} (p',p) =(4\pi)^2 \int_0^\infty \, dr\, r^2 \, j_{l'}(p'r)
j_{l}(pr) V_{l' l}^{JS}(r) \,
\end{eqnarray}
Using the Bessel function expansion for small argument $ j_l(x) =
x^l/(2l+1)!!  [1 - x^2/2(2l+3)+ \dots ] $ we get a low momentum
expansion of the potential matrix elements. 
We keep up to total order
${\cal O} (p^4, p'^4 , p^2 p'^2)$ corresponding to $S$-, $P$- and
$D$-waves as well as S-D and P-F mixing parameters,
\begin{eqnarray}
v_{00}^{JS}(p',p) 
&=& 
 \widetilde{C}_{00}^{JS} 
+ C_{00}^{JS}(p^2+p'^2) 
+ D^1_{00}{}^{JS} (p^4+p'^4) \nonumber \\
&+& D^2_{00}{}^{JS} p^2 p'^ 2 
+ \cdots 
\nonumber \\ 
v_{11}^{JS}(p',p) 
&=& 
p p' C_{11}^{JS} 
+ p p' (p^2+p'^2) D_{11}^{JS} 
+ \cdots 
\nonumber \\
v_{22}^{JS}(p',p) 
&=& 
p^2 p'{}^2 D_{22}^{JS} 
+ \cdots 
\nonumber   \\ 
v_{20}^{JS}(p',p) 
&=& p'{}^2 C_{20}^{JS} 
+ p'{}^2 p^ 2  D^1_{20}{}^{JS} + p'{}^4 D^2_{20}{}^{JS} 
+ \dots \nonumber   \\
v_{31}^{JS}(p',p) 
&=& 
p'{}^3 p D_{31}^{JS} 
+ \cdots 
\label{eq:Cts}
\end{eqnarray}
We use the spectroscopic notation and normalization of
Ref.~\cite{Epelbaum:2004fk}. The numerical results are shown in
Table~\ref{tab:Counterterms}. We see, again, a magnification of errors
in the short range contribution due to the lowering of the energy from
$350{\rm MeV}$ to $125 {\rm MeV}$ and a confirmation of the
universality between OPE and $\chi$TPE unveiled in
Ref.~\cite{Perez:2014kpa}. Actually we found a correlation pattern
which qualifies these counter-terms as good fitting parameters,
i.e. small statistical dependence and scheme independence.

\begin{table}
 \caption{\label{tab:Skyrme} Moshinsky-Skyrme parameters. We separate
the contribution from the delta-shells short range parameters
(corresponding to $r < r_c=1.8$fm) and the $\chi$TPE potential tail
(corresponding to $r> r_c$) . Units are: $t_0$ in ${\rm MeV} {\rm
fm}^3$, $t_1,t_2,W_0,t_U,t_T$ in ${\rm MeV} {\rm fm}^5$, and
$x_0,x_1,x_2$ are dimensionless.}
 \begin{ruledtabular}
 \begin{tabular*}{\columnwidth}{@{\extracolsep{\fill}} l D{.}{.}{4.5} 
 D{.}{.}{5.7}  D{.}{.}{5.5}  }
       & \multicolumn{1}{c}{$r < r_c$} 
       & \multicolumn{1}{c}{$r > r_c$} 
       & \multicolumn{1}{c}{Full} \\
 \hline \noalign{\smallskip}
$t_0$ &  -87.8(729)   & -382.3(91)   & -470.1(767)\\
$x_0$ &    -4.5(40)   &   -0.088(2)  &   -0.92(23)\\
$t_1$ &    77.2(98)   &  821.0(57)   &  898.1(117)\\
$x_1$ &    -1.2(1)    &   -0.00832(6)&   -0.11(1) \\
$t_2$ &   243.1(195)  & 2212.5(159)  & 2455.6(113)\\
$x_2$ &    -0.58(4)   &   -0.911(2)  &   -0.877(3)\\
$W_0$ &   105.8(30)   &    4.7       &  110.5(30) \\
$t_U$ &   148.1(49)   & 1132.8(30)   & 1281.0(56) \\
$t_T$ &  -569.4(301)  &-3836.0(90)   &-4405.4(279)\\
 \end{tabular*}
 \end{ruledtabular}
\end{table}

\begin{table}
 \caption{\label{tab:Counterterms} Potential integrals in different
partial waves. We separate the contribution from the delta-shells short
range parameters (corresponding to $r < r_c=1.8$fm) and the $\chi$TPE
potential tail (corresponding to $r> r_c$) . Units are:
$\widetilde{C}$'s are in $10^4 {\rm GeV}^{-2}$, $C$'s are in $10^4 {\rm
GeV}^{-4}$ and $D$'s are in $10^4 {\rm GeV}^{-6}$.}
 \begin{ruledtabular}
 \begin{tabular*}{\columnwidth}{@{\extracolsep{\fill}} l D{.}{.}{2.6} 
 D{.}{.}{4.5}  D{.}{.}{4.6}  }
       & \multicolumn{1}{c}{$r < r_c$} 
       & \multicolumn{1}{c}{$r > r_c$} 
       & \multicolumn{1}{c}{Full} \\
 \hline \noalign{\smallskip}
$\widetilde{C}_{^1S_0}$   & -0.079(14) &   -0.068(1) &   -0.15(1)   \\
            $C_{^1S_0}$   &  0.73(8)   &    3.48(2)  &    4.20(8)   \\
            $D_{^1S_0}^1$ & -6.2(17)   & -440.5(8)   & -446.7(19)   \\
            $D_{^1S_0}^2$ & -1.9(5)    & -132.1(2)   & -134.0(6)    \\
$\widetilde{C}_{^3S_1}$   &  0.051(18) &   -0.057(1) &   -0.006(19) \\
            $C_{^3S_1}$   & -0.078(28) &    3.42(2)  &    3.34(4)   \\
            $D_{^3S_1}^1$ &  0.12(4)   & -503.7(8)   & -503.6(8)    \\
            $D_{^1S_0}^2$ &  0.036(13) & -151.1(2)   & -151.1(2)    \\
            $C_{^1P_1}$   &  0.54(5)   &    5.92(4)  &    6.45(3)   \\
            $D_{^1P_1}$   & -2.0(2)    & -588.8(9)   & -590.8(8)    \\
            $C_{^3P_1}$   &  0.79(3)   &    2.934(8) &    3.72(3)   \\
            $D_{^3P_1}$   & -3.09(9)   & -246.1(3)   & -249.2(4)    \\
            $C_{^3P_0}$   &  0.046(12) &   -4.98(2)  &   -4.94(1)   \\
            $D_{^3P_0}$   &  0.55(32)  &  343.7(6)   &  344.2(5)    \\
            $C_{^3P_2}$   & -0.221(5)  &   -0.265(9) &   -0.486(8)  \\
            $D_{^3P_2}$   &  0.82(2)   &   -9.6(4)   &   -8.8(4)    \\
            $D_{^1D_2}$   & -0.44(5)   &  -70.5(1)   &  -70.9(1)    \\
            $D_{^3D_2}$   & -2.4(1)    & -363.0(2)   & -365.4(2)    \\
            $D_{^3D_1}$   &  2.2(3)    &  202.3(2)   &  204.5(3)    \\
            $D_{^3D_3}$   &  0.80(11)  &   -0.13(14) &    0.67(16)  \\
       $C_{\epsilon_1}$   & -1.13(6)   &   -7.60(2)  &   -8.72(6)   \\
       $D_{\epsilon_1}^1$ &  9.1(9)    &  998.9(5)   & 1008.0(8)    \\
       $D_{\epsilon_1}^2$ &  3.9(4)    &  428.1(2)   &  432.0(3)    \\
       $D_{\epsilon_2}$   &  0.59(7)   &   82.59(5)  &   83.18(6)   \\
 \end{tabular*}
 \end{ruledtabular}
\end{table}

\begin{figure*}
\centering
\includegraphics[width=\linewidth]{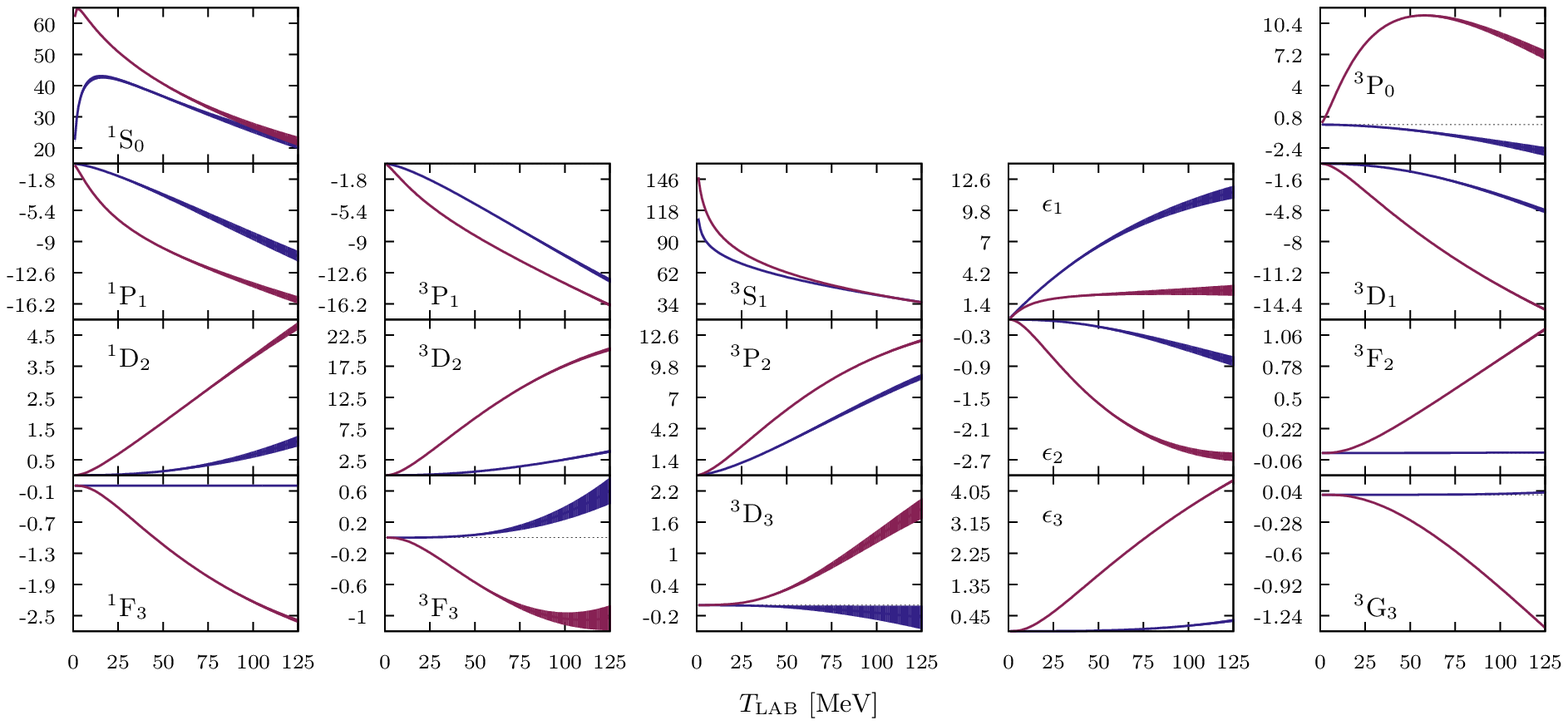}
\caption{(Color online) Short distance phase-shifts with statistical
  error bands in degrees (blue bands) for the $\delta$-shell
  $\chi$TPE-potential fitted up to a maximum ${\rm E_{\rm
      LAB}}=125{\rm MeV}$ obtained by eliminating all contributions of
  the potential with $r \ge 1.8 {\rm fm}$ and keeping just the
  Delta-shells potential (see main text). The complete phase-shifts
  (red bands) are drawn for comparison.}
\label{fig:decPhases}       % Give a unique label
\end{figure*}

\subsection{Short distance phase-shifts}

A complementary way to visualize the short distance structure of the
theory is by looking at the corresponding phase-shifts, $\delta^{\rm
  Short}_l$, which are those corresponding {\it just} to the short
distance part of the potential $V_{\rm DS}(r)$ in
Eq.~(\ref{eq:potential}). Because $V_{\rm DS}(r)$ has a range of
$r_c=1.8 {\rm fm}$ the partial wave expansion will converge for
$l_{\rm max} = p r_c $, and so we expect $\delta^{\rm Short}_{l_{\rm
    max}+1} \simeq 0 $ within the theoretical uncertainties. In our
case $l_{\rm max} = 2 $ which corresponds to D-waves, and so we expect
F, G and higher waves to produce negligible phase-shifts from the
short distance piece of the potential $V_{\rm DS}(r)$ in
Eq.~(\ref{eq:potential}). This is illustrated in
Fig.~\ref{fig:decPhases}, where F and G phase-shifts are very small
except for the marginal $^3F_3$ wave around $\simeq 100$ MeV.

We note that similar findings have been pointed out in the
deconstruction analysis of Ref.~\cite{Birse:2003nz} based on the
Nijmegen partial wave analysis. Our present study includes more and
coupled channels and takes statistical errors into account.

\subsection{Cut-off dependence}
\label{subsec:cutoff}

Our delta-shell potential can be interpreted as a UV regulator, where
$\Delta r \sim 1/\Lambda$ is the resolution scale and the short
distance cut-off $r_c$ the scale below which the interaction is
unknown~\footnote{These two scales are essentially identified in
  momentum space with the CM-momentum cut-off $\Lambda$ which is taken
  to be much larger than the branch cut singularity corresponding to
  $n$-pion exchange. One important reason for this is to avoid a
  cut-off quenching of the potential at long distances.}.  It is
interesting to consider the short distance cut-off dependence.
Actually we can try to replace the outermost delta-shells by the TPE
potential, that is, making $r_c=1.2$fm and therefore a further
reduction of parameters would take place. Moreover, restricting the
fit to lower energies corresponds to have about a twice larger
shortest de Broglie wavelength. Thus, one may naturally hope to be
more blind to the nucleon finite size effects.

We want to analyze here four possible effects, namely the
modifications due to i) taking fixed (and $\pi N$ motivated) chiral
constants vs NN fitted, ii) the counterterm structure, iii) the change
in the maximum fitted energy and iv) reduction of the short distance
cut-off.

\begin{table}
  \caption{\label{tab:Cut-offdependence} Different delta-shell
    potentials with chiral two pion exchange tail fitted to the
    self-consistent Granada database of~\cite{Perez:2013jpa}. The
    first line corresponds to the potential presented
    in~\cite{Perez:2013oba} and the second line to the potential of
    this work. The chiral constants of the fourth line were taken from
    a Ref.~\cite{Ekstrom:2013kea, Ekstrom:2014dxa} and used as fixed
    values during the $\chi^2$ minimization with respect of the
    delta-shell parameters. Highest counterterm column indicates the
    maximum angular momentum where at least one delta-shell strength
    coefficient is non-vanishing.}
  \begin{ruledtabular}
    \begin{tabular*}{\columnwidth}{@{\extracolsep{\fill}} l c D{.}{.}{3.6} 
        D{.}{.}{2.5} D{.}{.}{1.4}  c c}
      Max $T_{\rm LAB}$ & $r_c$ & 
      \multicolumn{1}{c}{$c_1$} & 
      \multicolumn{1}{c}{$c_3$} & 
      \multicolumn{1}{c}{$c_4$} & Highest  & $\chi^2/\nu$ \\
      MeV & fm & 
      \multicolumn{1}{c}{GeV$^{-1}$} & 
      \multicolumn{1}{c}{GeV$^{-1}$} & 
      \multicolumn{1}{c}{GeV$^{-1}$} & counterterm & \\
      \hline \noalign{\smallskip}
      350 & 1.8 & -0.4(11) & -4.7(6)  & 4.3(2) & $F$ & 1.08 \\
      350 & 1.2 & -9.8(2)  &  0.3(1)  & 2.84(5)& $F$ & 1.26 \\
      125 & 1.8 & -0.3(29) & -5.8(16) & 4.2(7) & $D$ & 1.03 \\
      125 & 1.2 & -0.92    & -3.89    & 4.31   & $P$ & 1.70 \\
      125 & 1.2 & -14.9(6) & 2.7(2)   & 3.51(9)& $P$ & 1.05 \\
    \end{tabular*}
  \end{ruledtabular}
\end{table}

The results are summarized in table~\ref{tab:Cut-offdependence} and we
proceed to discuss them. For instance, when we return to our
DS-$\chi$TPE fit up to $T_{\rm lab} = 350$MeV, we find that this
cut-off reduction from $r_c=1.8$fm to $r_c = 1.2$fm produces after
refitting parameters $\chi^2/N \sim 1.26$ with chiral constants
$c_{1,3,4}= -9.77, 0.31, 2.84$Gev$^{-1}$. Notice the unnaturally large
value for $c_1$ and how $c_3$ has positive value while most other
determinations from $\pi$N and NN give a negative value. We view these
features as a manifestation of the finite size effects of the nucleon
which become visible at the energy range extending up to pion
production threshold. A recent analysis of a Minimally non-local (and
continuous, i.e. plotable) nucleon-nucleon potentials with chiral
two-pion exchange including $\Delta$'s fully agrees with this
finding~\cite{Piarulli:2014bda}, namely a similar increase of the
$\chi^2/\nu = 1.3$, with a family of short distance potentials which
operate around $1.1 {\rm fm}$.

We now discuss the influence of the counterterm structure, namely
which partial waves are parameterized with the delta-shells
interpreted as an UV regulator. Thus, we may remove the $r \le r_c$
contribution to a given counterterm by setting {\it all} delta-shell
strength coefficients to zero in the corresponding partial
wave~\footnote{Actually this could eventually be done in such a way
  that {\it just} the advocated counterterms in a given power counting
  are included. We will not pursue this endeavour here and hope to do
  it in the future. This might require a comparative study on the
  admissible tolerance of a given power counting violation based both
  on statistical and systematic effects.}. If we take now the fixed
chiral constants of Ref.~\cite{Ekstrom:2013kea,Ekstrom:2014dxa}, and
assume the same counterterms structure on each partial wave and fit to
our database up to $T_{\rm LAB} = 125$Mev with our delta-shell inner
potential and $r_c=1.2$fm we get a statistically significantly larger
$\chi^2/N=1.7$. With this same structure one can substantially reduce
this value down to $\chi^2/N=1.05$ when the chiral contants are
allowed to vary. Their values $c_{1,3,4}=-14.87,2,71,3.51$GeV$^{-1}$
are again unnaturally large for $c_1$ and with the wrong sign for
$c_3$.

\subsection{Discussion}

We see that a feature of our calculation is that a fit up to $T_{\rm
  LAB}=125 {\rm MeV}$ fulfilling a good $\chi^2/\nu=1.02$ and passing
the normality test requires in addition to the $\chi$TPE potential
non-vanishing short distance contributions for S, P and D waves,
$\delta^{\rm Short}_l$.  As shown, a way of reducing short distance
D-wave phase-shifts is by reducing the value of $l_{\rm max}= p_{\rm
  max} r_c $ to $\sim 1$. This can be achieved either by reducing
$r_c$ below $1.8\, {\rm fm}$ or $p_{\rm max}$ or both.  For instance,
choosing $p_{\rm max}= m_\pi$ would correspond to $E_{\rm LAB} \le 40
{\rm MeV}$. Alternatively, one may choose $r_c=1 {\rm fm}$ and keep
$E_{\rm LAB} \le 125 {\rm MeV}$. According to our findings in
subsection \ref{subsec:cutoff} and our discussion on the anatomy of
the $NN$-potential~\cite{Perez:2013cza} taking $r_c \lesssim 1.8 {\rm
  fm}$ the nucleon size and quark exchange effects start playing a
role and the elementary particle assumption , upon which our
NN-potential approach is based, becomes debatable.

\section{Comparison with other low energy chiral potentials}
\label{sec:comparison}

This work introduces a new phenomenological Nucleon-Nucleon chiral
two-pion exchange potential fitted to pp and np scattering data up to
a laboratory energy of $125$MeV similar in spirit to other recent low
energy chiral
interactions~\cite{Ekstrom:2013kea,Gezerlis:2013ipa,Ekstrom:2014dxa,
  Gezerlis:2014zia} which have become popular in nuclear structure
calculations. We comment now on both approaches and the major
differences with ours from a statistical point of view.

\subsection{Momentum space optimized chiral potential at NNLO}

The momentum space self-denominated optimized chiral nucleon-nucleon
interaction at next-to-next-to-leading order
potential~\cite{Ekstrom:2013kea} provides a moderately acceptable
$\chi^2/\nu= 1.16$ value. It is based on the 1999 update of the
Nijmegen~\cite{Stoks:1993tb,Stoks:1994wp} database done with the
event of the CD Bonn potential analysis~\cite{Machleidt:2000ge}
with some minor modifications. With $\nu=N-P=1945-24$ degrees of
freedom, one should expect within $68\%$ confidence level a value
$\chi^2/\nu = 1 \pm \sqrt{2\nu}= 1 \pm 0.03$, which is excluded by
$5\sigma$~\footnote{Some pp data are excluded in the analysis
  of~\cite{Ekstrom:2013kea} on the basis of their extremely high
  precision which makes the $\chi^2$ value intolerably high. In our
  case these data are fully included in our $3\sigma$ self-consistent
  database, as we have no obvious reason to discard them.}.  In the
standard statistical jargon this means that the there is probability
$\sim 10^{-7}$ of erring when saying that the distribution {\it does
  not} obey a $\chi^2$ distribution. As we have stressed in our
previous works\cite{Perez:2014yla,Perez:2014kpa}, one may re-scale a
too large $\chi^2$ by a Birge factor to a new $\bar \chi^ 2=
(\chi^2/\chi^2_{\rm min}) \nu $, which by definition fulfills $\bar
\chi^2_{\rm min}/\nu=1$, {\it provided} the residuals of the fit are
normally distributed. In this case, a re-scaling of experimental
uncertainties, namely $\Delta O_i^{\rm exp} \to \sqrt{\chi^2_{\rm
    min}/\nu} \Delta O_i^{\rm exp} = \sqrt{1.16} \Delta O_i^{\rm exp}
$, would correspond to a bearable $7\%$ uncertainty in the error (the
error of the error). This is the kind of situation (too large
$\chi^2/\nu$) where the check for normality would be most
useful.~\footnote{Actually, that was the situation we encountered in
  the $\chi$TPE analysis up to a maximum energy of $E_{\rm LAB}=350
  {\rm MeV}$. }

\begin{figure}
\centering
\includegraphics[width=\linewidth]{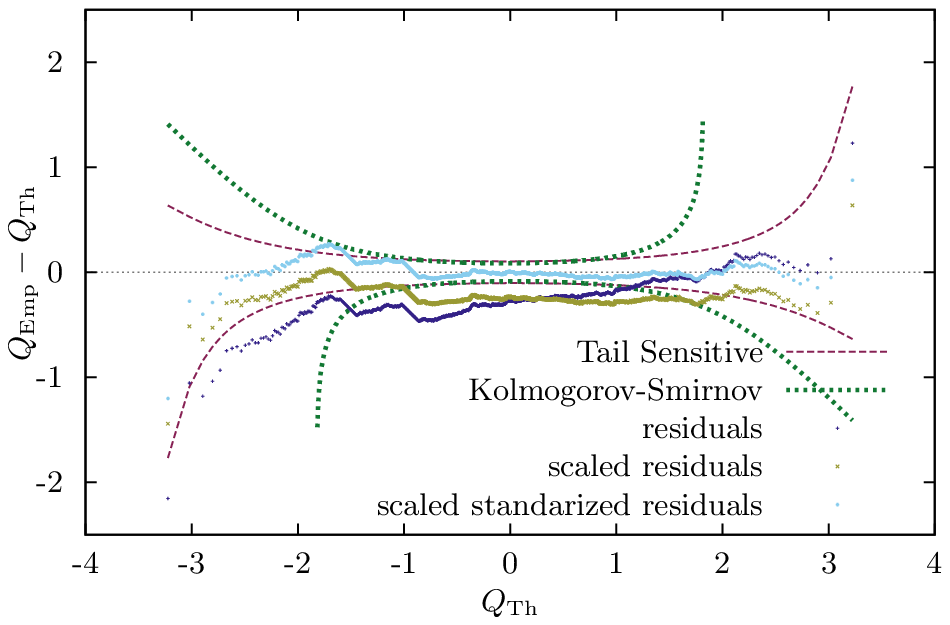}% figure caption is below the figure
\caption{(Color online) Rotated quantile-quantile plot for the
  residuals of the low energy optimized momentum space chiral
  potential of Ref.~\cite{Ekstrom:2014dxa} (dark blue crosses). The
  scaled (yellow crosses) and standardized (light blue crosses) are
  also shown. The Kolmogorov-Smirnov (dotted green line) and Tail
  sensitive (dashed red line) confidence bands with an $\alpha=0.05$
  significance level are also included.}
\label{fig:qqplot-pounders}       % Give a unique label
\end{figure}

In Ref.~\cite{Ekstrom:2014dxa} the reported skewness and excess
kurtosis of the histogram of residuals are $\Delta \mu_3'=0.06$ and
$\Delta \mu_4'=0.37$, respectively. The latter value is a bit too
high. Indeed, within a $68\%$ ($1\sigma$) confidence level we should
have $\Delta \mu_3' = \sqrt{17/\nu}$ and $\Delta \mu_4' =
4\sqrt{6/\nu}$, i.e.  $ \Delta \mu_3' = 0.09 $ and $ \Delta \mu_4' =
0.22$ respectively. The lack of normality is better unveiled in terms
of their QQ-plot.  They show a line resembling, $Q_{\rm Emp} \sim 1.35
Q_{\rm Th}$ ~\cite{Ekstrom:2014dxa} instead of the expected $Q_{\rm
  Emp} \sim Q_{\rm Th}$ straight line.  To better compare with our
Fig.~\ref{fig:qqplot}, that situation is recreated in a rotated QQ-plot
in Fig.~\ref{fig:qqplot-pounders} where the confidence bands are
adapted to the features of Ref.~\cite{Ekstrom:2014dxa}.~\footnote{The
  fact that there appear no points beyond the $Q_{\rm Th} > 3$ and
  $Q_{\rm Th} < -3$ is due to a truncation in the results shown in
  Ref.~\cite{Ekstrom:2014dxa}. The total $\chi^2$ obtained for the
  plot should be $(1848_{\rm data} + 108_{\rm normalizations}) \times
  1.16 = 2268.96$ while we get $2109.82$ for the about 1586 data.  A
  more quantitative analysis computing the $p-$value would require to
  totality of data or a truncated gaussian analysis, but will not
  change the main conclusions from the
  Fig.~\ref{fig:qqplot-pounders}.}

Since in Refs.~\cite{Ekstrom:2013kea,Ekstrom:2014dxa} a different
database from our $3\sigma$ self-consistent one was adopted, one may
think that re-doing the analysis might improve the normality
properties of the fit. This is unlikely, because normality requires
enough flexibility of the theory to encompass the fitted data up to
statistical fluctuations which are tolerated only thanks to the finite
number of data. Increasing the database should naturally decrease the
fluctuations. In our case, we have $N=925_{pp}+1743_{np}= 2688$ data
including normalizations vs the $N=1945$ data used in
Refs.~\cite{Ekstrom:2013kea,Ekstrom:2014dxa}. It is unlikely that the
bias introduced in the analysis of
Refs.~\cite{Ekstrom:2013kea,Ekstrom:2014dxa} will be compensated by
{\it adding} about 700 extra data. Thus, we attribute the lack of
normality to a lack of flexibility in the proposed interaction. The
question whether our self-consistent database is itself biased by our
own analysis is a pertinent one, but this could only be answered by
re-doing a data selection anew from scratch. Such an independent data
selection analysis would be most welcome to sort out these issues.

%Ndatum = Nobs + Nν − Nf loat = 1945 terms in the objective function,
%and Nν − Nf loat = 97 of them come from the normalization of certain
%data groups. The total number of parameters used in the optimization
%is Npar = Nf loat + NN N LO = 24,

\subsection{Local chiral potential}

\begin{figure}[htbp]
\begin{center}
\epsfig{figure=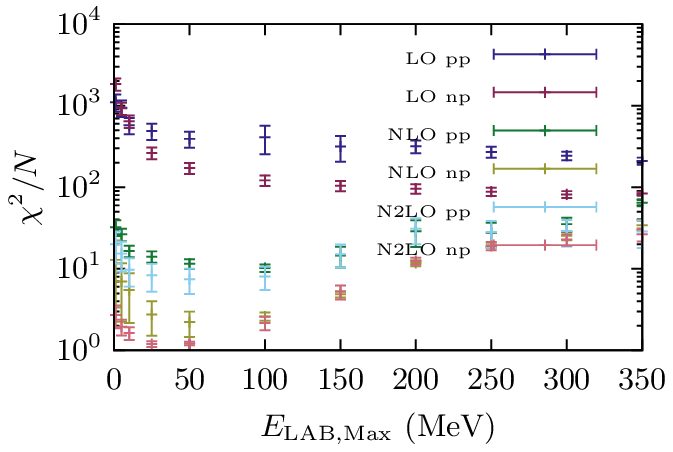,width=\linewidth} 
\end{center}
\caption{ (Color online) Values of $\chi^2/N$ for the local chiral
  potentials of Ref.~\cite{Gezerlis:2014zia} as a function of the
  maximum LAB energy (in MeV). We distinguish the different
  contributions building their LO,NLO and NNLO for np and pp.}
\label{fig:chi2-local-chiral}
\end{figure}

The local chiral potentials~\cite{Gezerlis:2013ipa,Gezerlis:2014zia}
fit phase shifts or low energy parameters in the lowest partial waves
taken as independent data and provide a sequence of LO,NLO and NNLO
schemes. An important feature of this potential concerns the regulator
which corresponds to a short distance potential of a range about
$1-1.2 {\rm fm}$.  We have implemented this potential and checked that
their phase-shifts are reproduced for all schemes. We can thus
confront this potential to the np and pp database and compute the
total $\chi^2$ as a function of the maximal LAB energy. The result is
shown in Fig.~\ref{fig:chi2-local-chiral} and as we see the smallest
value we get is $\chi^2/N \gtrsim 1$ for $T_{\rm LAB} \sim 40 {\rm
  MeV}$.  Nonetheless, our experience in comparing
phase-shift with PWA fits suggests that much better values could be
achieved with relatively small parameter changes. This is possibly an
effect due to the correlations among phase-shifts which in
Ref.~\cite{Gezerlis:2013ipa,Gezerlis:2014zia} are certainly
ignored. Given their wide applicability in nuclear structure
calculations, it would be interesting to perform a full PWA of these
local chiral potentials and test their normality.

\subsection{Discussion}

While the features of both the momentum space and the coordinate space
treatments discussed above are different regarding their
implementation details and statistical behavior, they share with our
present analysis the {\it same} chiral potential~\cite{Kaiser:1997mw}
at long distances $r \gtrsim 2 {\rm fm}$. The main difference relies
in the way the short distance pieces of the interaction are
represented. As we have mentioned Chiral perturbation theory
($\chi$pt) provides a power counting scheme to systematically include
pion exchange interactions in the complete NN
potential~\cite{Weinberg:1990rz, Ordonez:1992xp}. The several
phenomenological chiral potentials which have appeared in the
literature include the $\chi$pt derived pion exchange for the long
range part of the interaction and use counter-terms to describe the
\emph{unknown} short range part~\cite{Rentmeester:1999vw,
  Rentmeester:2003mf, Epelbaum:2005pn, Entem:2003ft, Ekstrom:2013kea,
  Gezerlis:2014zia}. However, most of these potentials give a fairly
large $\chi^2/\nu$ value when comparing with experimental data. In the
case with a desirable value $\chi^2/\nu \sim 1$ fitted to data
up to $125$ MeV~\cite{Ekstrom:2013kea}, the resulting residuals do not
follow the standard normal distribution~\cite{Ekstrom:2014dxa}. This
lack of normal residuals could indicate the presence of systematic
uncertainties.  Since the requirement of normally distributed data is
the basic building block of any least squares fit, any resulting
theory failing to fulfill this normality condition cannot be trusted
as a faithful representation of the scattering data and no reliable
propagation of statistical errors can be made. This does not rule out
to use them for nuclear structure calculations as it has been done in
the past where the normality and error propagation were never
addressed. Given this situation, it would be necessary and useful to
develop some understanding on how some conservative error estimates
could be done when normality is not fulfilled.

However, the main distinct feature which we see is the {\it necessity
  of a short distance D-wave component} when we fit up to $E_{\rm LAB}
= 125 {\rm MeV}$, a feature lacking recent low energy chiral
interactions~\cite{Ekstrom:2013kea,Gezerlis:2013ipa,Ekstrom:2014dxa,
  Gezerlis:2014zia}. We expect an improvement of normality and quality
including these additional terms in their analyzes. We remind that
according to the standard power counting invoked in those works the
N2LO chiral potential contains, in addition to the longer range OPE
and $\chi$TPE contributions, just S- and P-wave contact terms, while
the contact D-waves should have small contributions. This {\it a
  priori} condition is implemented in
Refs.\cite{Ekstrom:2013kea,Gezerlis:2013ipa,Ekstrom:2014dxa,
  Gezerlis:2014zia} by choosing a short distance regulator which has a
typical range of $r_c \sim 1 {\rm fm}$. According to our discussion
above on short distance phases, this is a way of killing the short
distance contribution to the $D-$waves, since $l_{\rm max} \sim 1$.
Our analysis shows instead a non-vanishing D-wave contribution from
the short distance piece of the interaction, within the statistical
uncertainties. This fact, while confirms the suitability of the
$\chi$TPE potential, suggests a more thorough revision of the standard
power counting assumed for the short distance components of the
interaction. This would require making a decision on setting an {\it a
  priori} acceptable $\chi^2/\nu$ value in order to declare
compatibility between a N2LO theory and the data used to extract the
corresponding counterterms.  As alreay mentioned, an adequate
assessment on the expected size of the power counting violations
requires an equally {\it a priori} estimate~\footnote{This is most
  clearly exemplified by the $\pi\pi$-scattering discussion of
  Ref.~\cite{Nieves:1999zb} on errors. Namely, if we have an expansion
  $O=O_0 + O_1 + \dots $ according to a declared power counting while
  $O_n$ and $\Delta O_n^{\rm stat}$ are the n-th order central value
  and statistical error of the observable, the condition for the
  expansion to be {\it predictive} and {\it convergent} at n-th order
  is $ \Delta O_n \ll O_{n+1} \ll O_n$. Implementing this analysis in
  the more complicated NN case requires a formidable effort which
  might be assisted by a Bayesian perspective. This would be along the
  lines of Ref.~\cite{Ledwig:2014cla} where the augmented $\chi^2$
  includes the natural size of counterterms as a the pertinent {\it
    prior} for minimization (see also
  e.g. Ref.~\cite{Furnstahl:2014xsa} for a recent proposal along these
  lines) and it is left for future research.}.  The discussion on the
specific power counting to be applied within $\chi$PT has been around
since the very beginning and most discussions have been carried out on
the basis of theoretical consistency~\cite{Nogga:2005hy,
  PavonValderrama:2005wv, Birse:2005um, Epelbaum:2006pt,
  Valderrama:2009ei, Valderrama:2011mv,Machleidt:2011zz}. Our finding
on the D-waves offers an excellent opportunity to discern on the basis
of experiment analysis among the several proposals on the market. As
already mentioned our analysis, while being completely satisfactory
from a statistical viewpoint it contains N2LO long range components in
the Weinberg power counting. Of course, it would be highly interesting
to extend the present investigation also to the N4LO chiral potential
proposed recently~\cite{Entem:2014msa}.

\section{Conclusions}
\label{sec:concl}

The use of chiral potentials in nuclear physics has become popular in
recent years as they are believed to incorporate essential low energy
QCD features. While this is formally correct a definite statement
supporting this expectation requires to make a decision on whether or
not the more than 8000 np and pp available data below pion production
threshold are described by the theory with a given confidence level.
So far the literature is lacking an estimate of the statistical
uncertainties propagated from low energy data only. Such analysis is
justified and performed with our new fit. A comparison with other high
energy error analyzes allows to evaluate the predictive power of low
energy chiral interactions and of course the statistical significance
of the included chiral effects.

We have taken the classical statistical point of view of validating
the theory using the least squares $\chi^2$-method. This method rests
on the first place on the assumption of normality of residuals, a
question which can be checked a posteriori and is not easy to
fulfill.  A lack of normality implies the presence of
systematic uncertainties in the analysis and excludes propagation of
statistical uncertainties. 

On the other hand, the available chiral potentials used in current
analyzes include OPE and TPE effects which limits the applicability of
the theory to about 100 MeV, which is the expected maximum relevant
energy for the binding of light nuclei. Thus, we have an interesting
opportunity to validate the chiral potential within a statistical
analysis of the corresponding low energy data within its range of
validity and usefulness for nuclear structure calculations. By using a
coarse grained delta-shell representation of the short distance
contribution, we observe a good description with an excellent fit to
scattering NN data and a quantitatively softer interaction as shown by
the Weinberg eigenvalues analysis. Special attention was given to
testing the normality of the residuals which allows to perform a sound
propagation of statistical errors. The assumption of normally
distributed experimental data was successfully tested. Statistical
error quantifications were made for potential parameters,
phase-shifts, scattering amplitudes, effective interaction parameters,
counter-terms and low energy threshold parameters. In all cases the
error bars were considerably larger than the full version of the
DS-$\chi$TPE potential fitted up to $T_{\rm Lab}=350$MeV. This fit
also allowed for a new determination of the chiral constants $c_1$,
$c_3$ and $c_4$ compatible with previous determinations from NN and
$\pi$N data.

Of course, reducing the fitted energy of the fit from 350 MeV to 125
MeV reduces the number of parameters but naturally increases the error
bars, not only in the extrapolated energy range, but also in the
active fitted range as we have about a third of np and pp scattering
data. This is so because the potential intertwines high and low energy
scattering data. We find, within uncertainties, unequivocal
non-vanishing short distance D-wave contributions to be essential both
for the fit and the normality behavior of the residuals. Thus, in
order to reduce the strength of the short distance D-wave pieces
without becoming sensitive to finite nucleon size details appears to
be to lower the maximum fitted energy. A comprehensive and systematic
study of such a maximum fitting energy dependence along the lines of
our previous study~\cite{NavarroPerez:2013iwa} but including normality
and uncertainties considerations would be possible and useful but
cumbersome and is left for future investigations.

An interesting follow up to this study would be the determination of a
low energy potential \emph{without} chiral components with the
corresponding statistical error analysis. A comparison of the
predictions given by both low energy interactions would show if the
inclusion of chiral effect is statistically significant or not,
shedding light into the actual predictive power of chiral interactions
determined by low energy data only. Some previous results have already
been advanced in Ref.~\cite{Amaro:2013zka} suggesting a lack of
significance of chiral interactions due to low energy uncertainty
enhancement and a more thorough study would be most desirable.

While we are not yet in a position to answer the question posed at the
beginning of this paper, we have provided and analyzed important
characteristics towards implementing suitable chiral interactions for
nuclear structure calculations. Along these lines the present work
shows that it is possible to fit NN scattering data with a Chiral Two
Pion Exchange potential fulfilling all necessary statistical
requirements up to 125 MeV inferring as a byproduct of the analysis
the short distance structure of the theory. At the same time if offers
a unique opportunity to discern, based on a direct comparison to
experimental scattering data, the possibility to establish the
validity of a power counting scheme which has the great advantage of
providing a priori error estimates. A complete study of this important
issue will be pursued elsewhere.

One of us (R.N.P.) thanks James Vary for hospitality at Ames. We thank
Maria Piarulli  and Rocco Schiavilla  for communications. This work is
supported by Spanish DGI (grant FIS2011-24149) and Junta de
Andaluc{\'{\i}a} (grant FQM225).  R.N.P. is supported by a Mexican
CONACYT grant.

%\section{Weinberg eigenvalues analysis}

%\bibliography{biblio}
%\bibliographystyle{utphys}

\end{document}